\documentclass[aps,prd,twocolumn,a4paper,superscriptaddress]{revtex4-1}
\usepackage{graphicx}
\usepackage{multirow}
\usepackage{latexsym}
\usepackage{hyperref}
\usepackage[british]{babel}
\usepackage{amsmath}
\begin{document}

\newcommand{\be}{\begin{equation}}
\newcommand{\ee}{\end{equation}}
\newcommand{\bq}{\begin{eqnarray}}
\newcommand{\eq}{\end{eqnarray}}
\newcommand{\bsq}{\begin{subequations}}
\newcommand{\esq}{\end{subequations}}
\newcommand{\bc}{\begin{center}}
\newcommand{\ec}{\end{center}}

\title{Collisions of cosmic strings with chiral currents}

\author{I. Yu. Rybak}
\email[]{Ivan.Rybak@astro.up.pt}
\affiliation{Centro de Astrof\'{\i}sica da Universidade do Porto, Rua das Estrelas, 4150-762 Porto, Portugal}
\affiliation{Instituto de Astrof\'{\i}sica e Ci\^encias do Espa\c co, CAUP, Rua das Estrelas, 4150-762 Porto, Portugal}
\affiliation{Faculdade de Ci\^encias, Universidade do Porto, Rua do Campo Alegre 687, 4169-007 Porto, Portugal}
\author{A. Avgoustidis}
\email[]{Anastasios.Avgoustidis@nottingham.ac.uk}
\affiliation{School of Physics and Astronomy, University of Nottingham, University Park, Nottingham NG7 2RD, England}
\author{C. J. A. P. Martins}
\email{Carlos.Martins@astro.up.pt}
\affiliation{Centro de Astrof\'{\i}sica da Universidade do Porto, Rua das Estrelas, 4150-762 Porto, Portugal}
\affiliation{Instituto de Astrof\'{\i}sica e Ci\^encias do Espa\c co, CAUP, Rua das Estrelas, 4150-762 Porto, Portugal}

\date{30 July 2018}

\begin{abstract}
We present an analytic study of cosmic superconducting chiral string collisions in Minkowski space, applying the kinematic constraints that arise from the relevant generalization of the Nambu-Goto action. In particular, we revisit the solution for chiral superconducting cosmic strings and demonstrate that Y junction production for such strings is possible. We consider the collision of chiral current-carrying straight strings and obtain the region in ``angle-velocity" space that allows the production of string junctions. This study contributes to the understanding of the complex evolution of chiral superconducting string networks.

\end{abstract}
\pacs{98.80.Cq, 11.27.+d, 98.80.Es}
\keywords{Cosmology; Cosmic strings; String collisions, chiral currents}
\maketitle

%%%%%%%%%%%%%%%%%%%%%%%%%%%%%%%%%%%%%%%%%%%%%%%%%%%%%%%%%%%%%%%%%%%%%%%%%%%%%%%%%%
\section{Introduction}

Cosmic string networks are theoretically predicted relics of early universe physics that can survive until the present day. More than four decades after their prediction by Tom Kibble \cite{Kibble}, cosmic strings still attract attention as a generic byproduct of a wide range of inflationary and high energy particle physics models: brane inflation \cite{BurgessMajumdarNolteQuevedoRajeshZhang, DvaliKalloshProeyen, DvaliVilenkin, PolchinskiCopelandMyers,SarangiTye,FirouzjahiTye,JonesStoicaTye},  supersymmetric grand unified theory \cite{JeannerotRocherSakellariadou,CuiMartinMorrisseyWells,JeannerotPostma, AchucarroCeli,DavisMajumdar,Allys}, and other models of inflation \cite{KoehnTrodden,LazaridesPeddieVamvasakis,ChernoffTye} and particle physics \cite{BRANDENBERGER, BRANDENBERGER2, KawasakiKen'ichiSekiguchi, GripaiosRandal-Williams}. The diversity of these models endows cosmic strings with different properties. Some of them, strings arising in models of the brane inflation type, can feature Y junction configurations produced by string collisions. Kinematic treatment via the Nambu-Goto action allowed the study of the dynamics of Y junctions and revealed the conditions that the relative orientation and velocity of the colliding strings must satisfy in order to create a junction \cite{CopelandKibbleSteer, CopelandKibbleSteer2, CopelandFirouzjahiKibbleSteer}. Analytic studies were later confirmed by numerical field theory simulations \cite{SalmiAchucarroCopelandKibblePutterSteer, BevisSaffin}.

At the same time, a number of models suggest the presence of non-trivial internal structure on cosmic strings. This can be caused by a coupling between the field forming the cosmic string and other fields, by trapped charged fermion modes along the string~\cite{Witten} (which is common in supersymmetric models~\cite{DavisDavisTrodden,DavisDavisTrodden2}), by trapped vector fluxes on non-Abelian strings~\cite{Everett}, and other specific mechanisms (for example breaking of an accidental symmetry in SU(2) strings~\cite{HindmarshRummukainenWeir}). The first example of an effective description of such charged strings was done for superconducting strings in \cite{Witten}. This subsequently led to the appearance of a number of different effective models aiming to capture the diversity of string properties by effective worldsheet currents \cite{CARTER1989, Peter1992, CarterPeter, CarterPeter2, Carter2000, Carter90, Vilenkin, Carter95}.

We anticipate that Y junction production can also arise more generally for strings with non-trivial structure. A comprehensive study of such a situation was presented in \cite{SteerLilleyYamauchiHiramatsu}. Specifically, the authors of \cite{SteerLilleyYamauchiHiramatsu} considered collisions of strings with \textit{electric} current (time-like state parameter $\kappa <0$ - see Section \ref{Solution in Minkowski space for strings with chiral currents} below for a more precise definition of $\kappa$) and \textit{magnetic} current (space-like state parameter $\kappa>0$). It was found that for such strings the collision cannot lead to the formation of Y junction, unless the newly formed string is described by a more general equation of state.

The current work aims at studying the kinematic constraints for colliding strings with chiral type of current ($\kappa=0$), which was not covered in \cite{SteerLilleyYamauchiHiramatsu}. We show that the junction production for such strings is possible. In particular, we quantify the influence of the chiral current on the dynamics of Y-junctions and ``angle-velocity" restrictions for Y junction production.

%%%%%%%%%%%%%%%%%%%%%%%%%%%%%%%%%%%%%%%%%%%%%%%%%%%%%%%%%%%%%%%%%%%%%%%%%%%%%%%%%%
\section{Solution in Minkowski space for strings with chiral currents} \label{Solution in Minkowski space for strings with chiral currents}

In this section we consider the superconducting string solution described by the general action \cite{RybakAvgoustidisMartins}
\begin{equation}
\label{Action}
S = - \mu_0 \int f(\kappa, \Delta/\gamma),
\end{equation}
where $f$ is an arbitrary function, $\mu_0$ is a constant defined by the symmetry breaking scale with dimensions $[\textit{Energy}]^2$ and
\begin{subequations}
\begin{align}
   \label{TermsA}
     \phi^{,a} \phi^{,b} \gamma_{ab} \equiv \kappa, \\ 
     \label{TermsB}
     \varepsilon^{ac} \varepsilon^{bd} \gamma_{ab} \gamma_{cd} \equiv \gamma, \\
     \label{TermsC} 
   \varepsilon^{ac} \varepsilon^{bd} \gamma_{ab} \phi_{,c} \phi_{,d} \equiv \Delta,   
\end{align}
\end{subequations}
with Levi-Civita symbol $\varepsilon^{ac}$, induced metric on the string worldsheet $\gamma_{ab}$ and (\ref{TermsA}) defining the state parameter.

The chiral condition corresponds to the case when
\begin{equation}
\label{ChiralCurrent}
\Delta, \, \kappa \rightarrow 0.
\end{equation}

The chiral condition (\ref{ChiralCurrent}) doesn't imply that the current disappears, it just means that the current is described by a null (light-like) vector. Let's obtain the equations of motion, and then take the limit (\ref{ChiralCurrent}). The equations of motion for the action~(\ref{Action}) can be written in the form
\begin{equation}
   \label{EqOfMotGeneral1}
     \partial_a \left( \mathcal{T}^{ab} x^{\mu}_{,b} \right) = 0,  
\end{equation}
\begin{equation}
   \label{EqOfMotGeneral2}
    \partial_a \left( \mathcal{S}^{ab} \phi_{,b}  \right) = 0,  
\end{equation}
with
\begin{equation}
   \label{Tpart}
    \mathcal{T}^{ab} \equiv \sqrt{-\gamma} \left( \gamma^{ab} f + \theta^{ab} \right),  
\end{equation}
where $\theta^{ab} \equiv 2 \left( \frac{\partial f}{\partial \kappa} \gamma^{ac} \gamma^{bd} + \frac{1}{\gamma}\frac{\partial f}{\partial \Delta} \varepsilon^{ac} \varepsilon^{bd} \right) \phi_{,c} \phi_{,d}$ and
\begin{equation}
   \label{Spart}
    \mathcal{S}^{ab} \equiv \sqrt{-\gamma} \left( \gamma^{ab} \frac{\partial f}{\partial \kappa} +\frac{1}{\gamma} \frac{\partial f}{\partial \Delta} \varepsilon^{ac} \varepsilon^{bd} \gamma_{cd} \right),  
\end{equation}
where the limit (\ref{ChiralCurrent}) has been taken.

To find an exact solution, we will follow the method developed in \cite{Blanco-PilladoOlumVilenkin}. To do so, we need to find a parametrisation of the worldsheet such that the usual (without current) equations of motion for the string are valid
\begin{equation}
   \label{EqOfMotGeneralFlat}
     \partial_a \left( \eta^{ab} x^{\mu}_{,b} \right) = 0,  
\end{equation}
which requires that
\begin{equation}
   \label{TEta}
     \mathcal{T}^{ab} = \eta^{ab},  
\end{equation}
where $\eta^{a b}$ is a $2$x$2$ Minkowski metric.
    
In order to show that it is possible to choose such a parametrization, to satisfy relations (\ref{EqOfMotGeneralFlat}) and (\ref{TEta}), we repeat the procedure from \cite{Blanco-PilladoOlumVilenkin}, i.e. study the $\mathcal{T}^{ab}$ determinant
\begin{equation}
\begin{gathered}
   \label{Det}
     \det \mathcal{T}^{ab} = -\gamma\,  \text{det} \left( f \gamma^{a b} + \theta^{a b}  \right) = \\
      = - \det \left( f \delta_{c}^{a} + \theta_{c}^{a} \right) = -f^2 - f \text{Tr} \theta_{c}^{a} - \det  \theta_{c}^{a}. 
\end{gathered}
\end{equation}

Note the absence of $\mu_0$ comparing with \cite{Blanco-PilladoOlumVilenkin}. If the expression (\ref{Det}) is equal to $\text{det} \eta^{ab} = -1$, there should exist a parametrization that allows to satisfy conditions (\ref{EqOfMotGeneralFlat}--\ref{TEta}). Recalling that $f \rightarrow 1$ for the chiral current, we see that the second and third terms in the Eq.(\ref{Det}) should be zero. Using the chiral condition (\ref{ChiralCurrent}) one can show that $\text{Tr}\theta^a_c=0$ and $\text{det}\theta^a_c=0$ (details can be found in \cite{Blanco-PilladoOlumVilenkin}). As a result, for chiral currents it is possible to choose a parametrization that leads to equations of motion (\ref{EqOfMotGeneralFlat}), the general solution of which is
\begin{equation}
   \label{StringSolution}
      \textbf{x} = \frac{1}{2} \left( \textbf{a}(\sigma+\tau) +  \textbf{b}(\sigma-\tau) \right), 
\end{equation}
where $\textbf{a}(\sigma+\tau)$ and $\textbf{b}(\sigma-\tau)$ are arbitrary vector-valued functions.

We still need to consider the equation for the field $\varphi$ (\ref{EqOfMotGeneral2}). Using the definition (\ref{Spart}), we rewrite equation (\ref{EqOfMotGeneral2}) explicitly
\begin{equation}
\begin{split}
   \label{EqOfMotPhi}
    & \qquad \partial_a \left( \sqrt{-\gamma} \left( \gamma^{ab} \frac{\partial f}{\partial \kappa} +\frac{1}{\gamma} \frac{\partial f}{\partial \Delta} \varepsilon^{ac} \varepsilon^{bd} \gamma_{cd} \right) \phi_{,b}  \right) =  \\
    & \qquad \; = \partial_a \left( \sqrt{-\gamma} \gamma^{ab} \left(  \frac{\partial f}{\partial \kappa} - \frac{\partial f}{\partial \Delta}  \right) \phi_{,b}  \right) =0.
\end{split}
\end{equation}

At the same time, it is seen that from equations (\ref{TEta}) and (\ref{Tpart}) we can obtain the following relations 
\begin{equation}
\begin{split}
   \label{Relations}
    & \qquad \; \sqrt{-\gamma} \gamma^{ab} = \mathcal{T}^{ab} - \sqrt{-\gamma} \theta^{ab}, \\
    & \sqrt{-\gamma} \varepsilon^{ac} \varepsilon^{bd} \gamma_{cd} = \varepsilon^{ac} \varepsilon^{bd} \left( \mathcal{T}_{cd} - \sqrt{-\gamma} \theta_{cd} \right).
\end{split}
\end{equation}

Let's use the calculated expressions (\ref{Relations}) in equation (\ref{EqOfMotPhi}). Recalling the proven identity (\ref{TEta}), we obtain the following construction
\begin{equation}
   \label{EqOfMotPhi20}
     \partial_a \left[ \left( \eta^{ab} - \sqrt{-\gamma} \theta^{ab} \right) \left( \frac{\partial f}{\partial \kappa} - \frac{\partial f}{\partial \Delta}  \right)  \phi_{,b} \right] =0.
\end{equation}

Simplifying the expression (\ref{EqOfMotPhi20}) we obtain the final equation for the field $\phi$ as
\begin{equation}
   \label{EqOfMotPhi2}  
     \partial_a \left[  \left( \frac{\partial f}{\partial \kappa} - \frac{\partial f}{\partial \Delta}  \right) \eta^{ab} \phi_{,b} \right] =0,    
\end{equation}

Using the condition $\theta^{ab} \phi_{,a} \phi_{,b}=0$ together with (\ref{Tpart}) and (\ref{TEta}), one can conclude that
\begin{equation}
   \label{EqOfMotPhi3}
     \eta^{ab} \phi_{,a} \phi_{,b} = 0.    
\end{equation}

Let's recall that $\frac{\partial f}{\partial \kappa}$ and $\frac{\partial f}{\partial \Delta}$ are constants for the chiral current. As a result, assuming that $\frac{\partial f}{\partial \kappa} - \frac{\partial f}{\partial \Delta} \neq 0 $, the general solution of equations (\ref{EqOfMotPhi2}), (\ref{EqOfMotPhi3}) is 
\begin{equation}
   \label{SolutionPhi}
     \phi = \frac{1}{2} F(\sigma \pm \tau),    
\end{equation}
where $\frac{1}{2}$ was chosen for convenience and in the later discussion we will use the $``+"$ sign, which does not reduce the generality of these considerations.

There are additional constraints that should be imposed. These conditions can be obtained by considering the metric $\gamma^{ab}$. The equation for the metric components that arises from the chiral condition is
\begin{equation}
\begin{split}
   \label{GammaEq}
    \gamma^{00}+\gamma^{11}+2\gamma^{01} = 0 \quad \text{or} \quad \gamma_{00} + \gamma_{11} - 2 \gamma_{01} = 0.
 \end{split}
\end{equation}

Using the definition $\gamma_{ab} = x_{,a}^{\mu} x_{\mu,b}$ for (\ref{GammaEq}), we obtain the equation
\begin{equation}
\begin{split}
   \label{Vector}
    & \dot{x}^{\mu} \dot{x}_{\mu} + x^{\mu \prime} x_{\mu}^{\prime} - 2\dot{x}^{\mu}  x_{\mu}^{\prime} = \left( \dot{x}^{\mu} - x^{\mu \prime}\right) \left( \dot{x}_{\mu} - x_{\mu}^{\prime}\right) = 0,
 \end{split}
\end{equation}
which allows us to conclude that
\begin{equation}
\begin{split}
   \label{VectorNorm}
    & |\textbf{b}^{\prime}|=1.
 \end{split}
\end{equation}

Using the identity $\theta^{ab} \theta_{bc} = 0$ and (\ref{TEta}) together with the fact that $\theta^{ab}$ and $\gamma^{ab}$ are symmetric, we can conclude that 
\begin{equation}
\begin{split}
   \label{AddEq}
    & \frac{1}{\sqrt{-\gamma}} \left( \gamma_{ab}  - \theta_{ab} \right)  = \eta_{ab}.
  \end{split}
\end{equation}

Calculating the \textit{``01"} components in equation (\ref{AddEq}), we conclude that
\begin{equation}
\begin{split}
   \label{UsefulQuant}
     \gamma_{01} = \frac{1}{4} \left( 1 - |\textbf{a}^{\prime}|^2 \right)
     \end{split}
\end{equation}
and with the condition (\ref{AddEq}), one can obtain the final connection between the solutions (\ref{StringSolution}) and (\ref{SolutionPhi})
\begin{equation}
   \label{ConditionPhi}
   1 - |\textbf{a}^{\prime}|^2 = 2 \left( \frac{\partial f}{\partial \kappa} - \frac{\partial f}{\partial \Delta} \right) F^{\prime 2}.
\end{equation}

One can show that components \textit{``00"} and \textit{``11"} for equation (\ref{AddEq}) coincide with equation (\ref{ConditionPhi}). As a result, we obtained the general solution for the superconducting string (\ref{StringSolution}), (\ref{SolutionPhi}) together with condition (\ref{ConditionPhi}) for any type of chiral current. The difference between the obtained solutions and previously studied particular cases \cite{CarterPeter2, DavisKibblePicklesSteer, Blanco-PilladoOlumVilenkin} appears only in the multiplier $\left( \frac{\partial f}{\partial \kappa} - \frac{\partial f}{\partial \Delta} \right)$ in the equation (\ref{ConditionPhi}). 

%%%%%%%%%%%%%%%%%%%%%%%%%%%%%%%%%%%%%%%%%%%%%%%%%%%%%%%%%%%%%%%%%%%%%%%%%%%%%%%%%%
\section{Junctions with chiral currents} \label{Junctions with chiral currents}

Let's consider the dynamics of junctions for current-carrying strings. For \textit{electric} (time-like current $\kappa<0$) and \textit{magnetic} (space-like current $\kappa>0$) regimes of currents on strings see \cite{SteerLilleyYamauchiHiramatsu}. Here we present the corresponding result for the chiral current $\kappa=0$, which has not been studied to date.

Let's start from the definition of the action for three connected strings \cite{Sharov1997, Hooft} with adaptation for the presence of a current $\phi$ \cite{SteerLilleyYamauchiHiramatsu} 
\begin{equation}
\begin{split}
\label{ActionJunctCurrent}
& \qquad S_{\text{all}} = \sum_i S_i \; \Theta \left( s_i(t) - \sigma_i \right) +\\
& + \sum _i \int dt \, \mathrm{f}_{\mu i} \left( x_i^{\mu} (s_i(t),t) - \mathcal{X}^{\mu}(t) \right) + \\
& \quad +  \sum_i \int dt \, \mathrm{g}_i \left( \phi_i (s_i(t),t) - \Phi(t) \right), 
\end{split}
\end{equation}
where $S_i$ refers to the form of the action (\ref{Action}), $\mathrm{g}_i$ is a Lagrange multiplier for the current, $\Phi$ defines the value of the current at the point where the three strings are connected and the index $i= 1,2,3$ denotes each of the three strings.

Varying the action (\ref{ActionJunctCurrent}) with respect to $x_i^{\mu}$ and $\phi_i$, we obtain the usual equations of motion (\ref{EqOfMotGeneral1}) and (\ref{EqOfMotGeneral2}). The boundary terms proportional to $\delta(s_i(t) - \sigma_i)$, using (\ref{TEta}) and (\ref{EqOfMotPhi2}), can be expressed as
\begin{equation}
\begin{split}
\label{BoundTerms}
& \, \, \mu_i \eta^{ab} x_{i,a}^{\mu} \lambda_{b} = \mathrm{f}_i^{\mu}, \\ 
&  \mu_i \mathcal{D}_i \eta^{ab} \phi_{,a} \lambda_{b} = \mathrm{g}_i,
\end{split}
\end{equation}
where $\lambda_a = \left\lbrace \dot{s}_i,\; -1  \right\rbrace$ and $\mathcal{D}_{i} = 2 \left(\frac{\partial f_i}{\partial \kappa_{i}} -  \frac{\partial f_i}{\partial \Delta_{i}} \right)$.

The variation of the action (\ref{ActionJunctCurrent}) with respect to $\mathcal{X}_i^{\mu}$ and $\Phi$ gives us 
\begin{equation}
\begin{split}
\label{VarXPhi}
& \sum_i \mathrm{f}_i^{\mu} = 0, \\ 
& \, \sum_i \mathrm{g}_i = 0.
\end{split}
\end{equation}

Finally, variation of the action (\ref{ActionJunctCurrent}) with respect to $\mathrm{f}_i^{\mu}$ and $\mathrm{g}_i$ provides us the following relations
\begin{equation}
\begin{split}
\label{Varfg}
& x_i^{\mu}(s_i(t),t) = X^{\mu}(t), \\ 
& \; \phi_i(s_i(t),t) = \Phi(t).
\end{split}
\end{equation}

It is seen that for the chiral current the situation is similar to the one considered in \cite{CopelandKibbleSteer, CopelandKibbleSteer2}, i.e. we can separate the string solution in ingoing and outgoing modes. As a result, the modes that move outwards from the junction are determined by the ingoing modes. However, the situation with the current is a bit more subtle. The current on a string propagates only in one direction. Additionally, while we know the properties of the colliding strings (in particular we know $\mathcal{D}$ for both strings) it is not clear how to define the value of $\mathcal{D}$ for the new string. Should it be a free parameter, or can it somehow be restricted by other arguments as it was done for the tension of superstrings \cite{WITTEN1996}? For now, we assume that $\mathcal{D}$ is a free parameter defined by kinematic constraints.

\subsection{Collisions of identical strings} \label{Collisions of identical strings}

\begin{figure}[ht]
\centering
		\includegraphics[width=8.0cm]{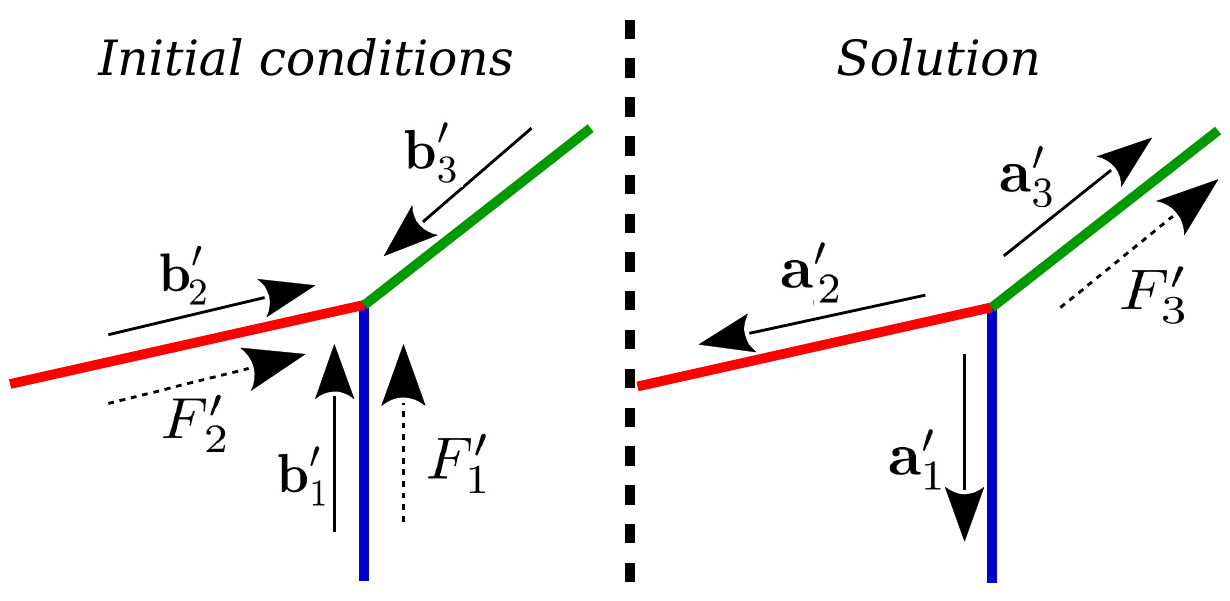}
\caption{Collision of strings with currents, whose moving modes are represented by $F^{\prime}_{i}$ and ingoing/outgoing modes of strings are shown by vectors $\textbf{a}_i^{\prime}$ and $\textbf{b}_i^{\prime}$ respectively.}
\label{fig:LRC}
\end{figure}

Let's consider the collision of two identical strings ($\mu_1=\mu_2$, $F_{1}^{\prime}=F_{2}^{\prime}$, $\mathcal{D}_1=\mathcal{D}_2$), assuming $\mathcal{D}_3$ is a free parameter. We can choose the parametrization of the colliding strings in the same way as it was done for strings without currents \cite{CopelandKibbleSteer, CopelandKibbleSteer2}, i.e. all vectors $\textbf{b}^{\prime}$ represent modes that are moving towards the junction, while $\textbf{a}^{\prime}$ are outgoing modes.
From (\ref{SolutionPhi}) it is seen that the choice of the sign of $\tau$ for the function $F$ defines the current direction on a string. Hence, we can choose whether the current propagates toward the vertex between strings or outwards. Without loss of generality, let's consider the collision of strings in which currents are moving towards the junction, see figure \ref{fig:LRC}. 

\begin{figure*}[ht!]
\centering
		\includegraphics[width=15.5cm]{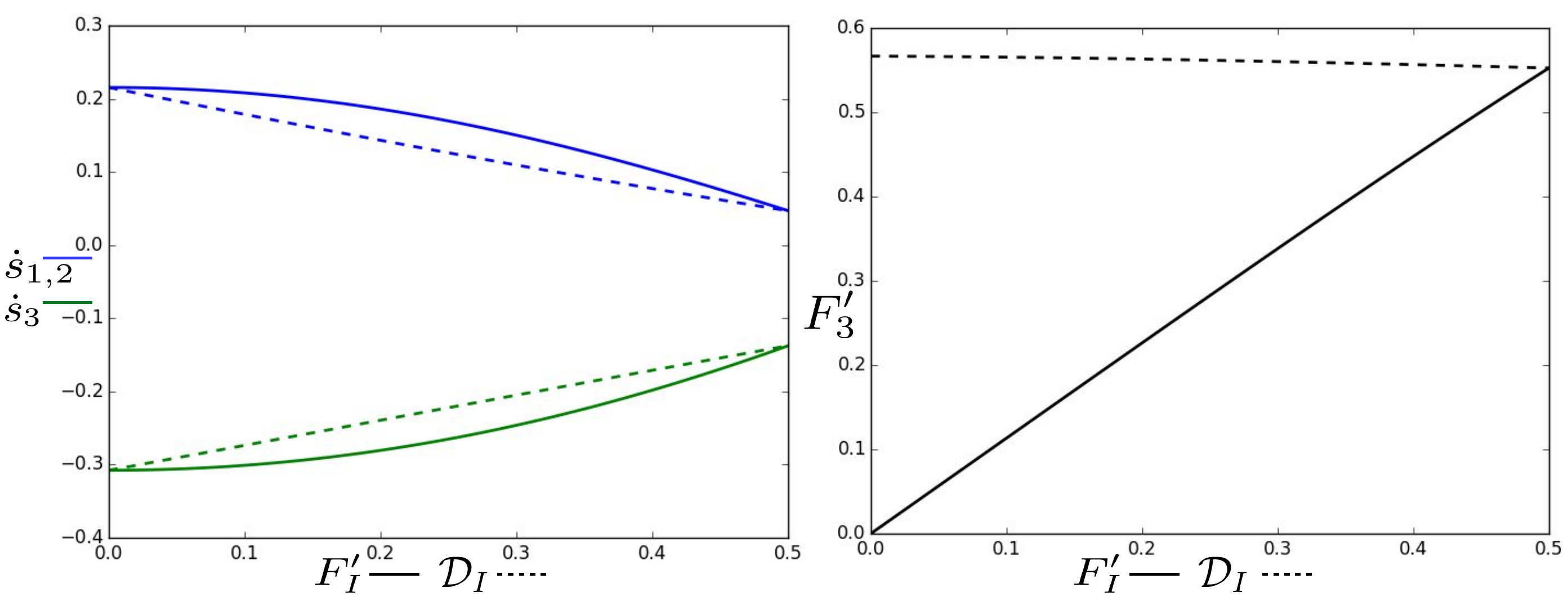}
\caption{The left panel shows how the growth/decrease rate $\dot{s}_i$ for the string configurations is changing due to variations of current properties, represented by $F_I^{\prime}$ and $\mathcal{D}_I$. The right panel shows the value of the current $F_3^{\prime}$ generated on the junction for different values of $F_I^{\prime}$ and $\mathcal{D}_I$. All solid lines represent variation of $F_I^{\prime}$, with fixed $\mathcal{D}_I=0.5$, while all dashed lines represent variation of $\mathcal{D}_I$ with fixed $F_I^{\prime}=0.5$. These calculations are carried out for string tensions $\mu_1=\mu_2=1$, $\mu_3=1.2$ when all vectors $\textbf{b}_i^{\prime}$ are orthogonal. }
\label{fig:S3F3Symm}
\end{figure*}

In this situation the string solution together with the current can be written as

\begin{equation}
\begin{split}
\label{Colliding}
& \textbf{x}_{1,2}(t,\sigma) = \frac{1}{2} \left( \textbf{a}_{1,2}(\sigma+t)+\textbf{b}_{1,2}(\sigma - t) \right), \\ 
& \qquad \quad \phi_{1,2}(t,\sigma) = \frac{1}{2} F_{1,2}(\sigma-t),
\end{split}
\end{equation}
\begin{equation}
\label{CollidingNorm}
\text{with} \qquad |\textbf{a}^{\prime}_{1,2}|^2=1, \quad |\textbf{b}_{1,2}^{\prime}|^2 = 1-\mathcal{D}_{1,2} |F^{\prime}_{1,2}|^2.
\end{equation}

Therefore, the third string ansatz has to be 
\begin{equation}
\begin{split}
\label{JunctionSol}
& \textbf{x}_{3}(t,\sigma) = \frac{1}{2} \left( \textbf{a}_{3}(\sigma+t)+\textbf{b}_{3}(\sigma - t) \right), \\ 
& \qquad \quad \phi_{3}(t,\sigma) = \frac{1}{2} F_{3}(\sigma+t),
\end{split}
\end{equation}
\begin{equation}
\label{JunctionSolNorm}
\text{with} \qquad |\textbf{b}^{\prime}_{3}|^2=1, \quad |\textbf{a}_{3}^{\prime}|^2 = 1-\mathcal{D}_3 |F^{\prime}_{3}|^2.
\end{equation}

Using solutions (\ref{Colliding}), (\ref{JunctionSol}) we can rewrite  (\ref{BoundTerms}) with (\ref{VarXPhi}) as
\begin{equation}
\begin{split}
\label{Sum1} 
& \; \, \sum_i \mu_i \left( \textbf{a}_i^{\prime} (1+\dot{s}_i) + \textbf{b}_i^{\prime} (1-\dot{s}_i) \right) = 0, \\ 
& \sum_{I} \mu_I \mathcal{D}_I F^{\prime}_I (\dot{s}_I-1) = \mu_3 \mathcal{D}_3 F^{\prime}_3 (\dot{s}_3+1).
\end{split}
\end{equation}

\begin{figure*}[ht]
\centering
		\includegraphics[width=15.5cm]{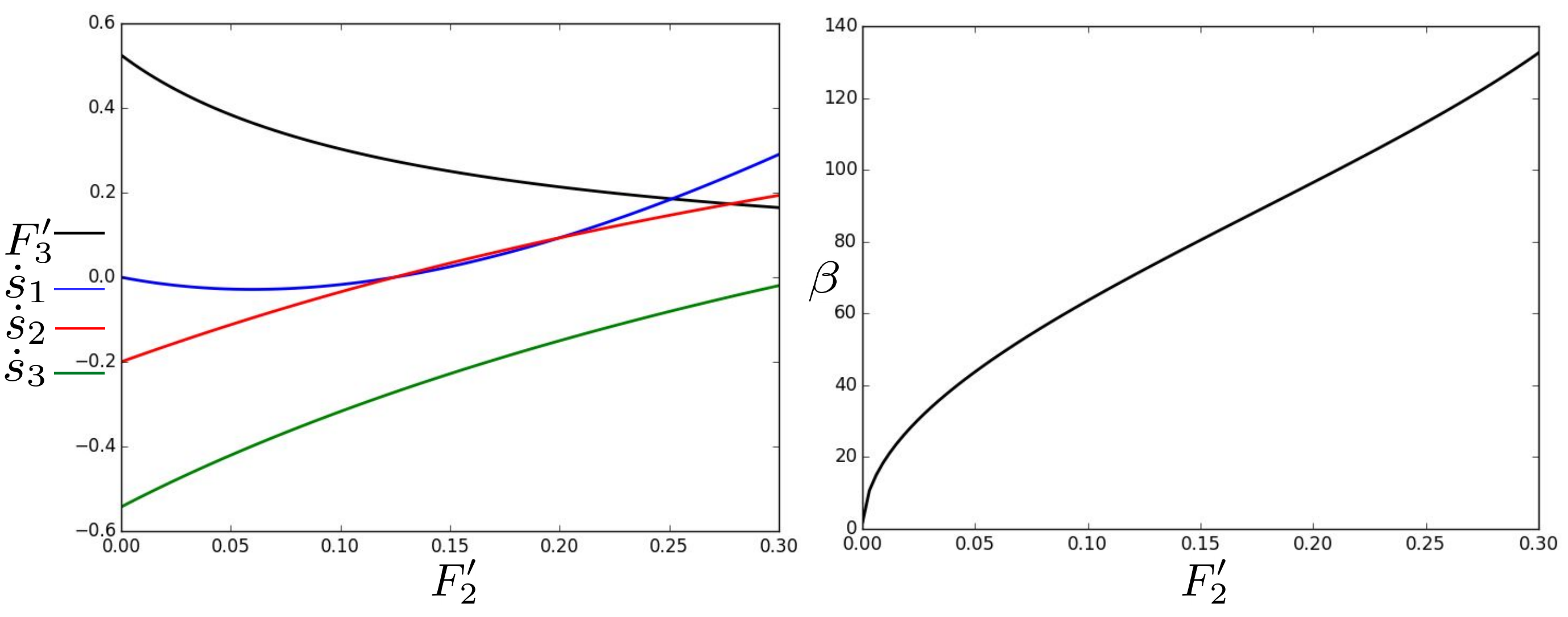}
\caption{The left panel shows how the growth/decrease rates $\dot{s}_i$ and the current $F_{3}^{\prime}$ for strings depend on $F_2^{\prime}$. The right panel shows the value of the angle $\beta$ between vectors $\textbf{b}_2$ and $\textbf{b}_3$. Calculations are carried out for string tensions $\mu_1=1$, $\mu_2=1.2$, $\mu_3=1.4$, while $F_1^{\prime}=0.2$, $\mathcal{D}_1=0.1$ and $\mathcal{D}_2=0.2$. The angles between vectors $\textbf{b}_1, \, \textbf{b}_2 $ and $ \textbf{b}_1, \, \textbf{b}_3 $ are free and were chosen $\pi/2$. }
\label{fig:S3F3NonSymm}
\end{figure*}

\begin{figure*}[ht]
\centering
		\includegraphics[width=15.5cm]{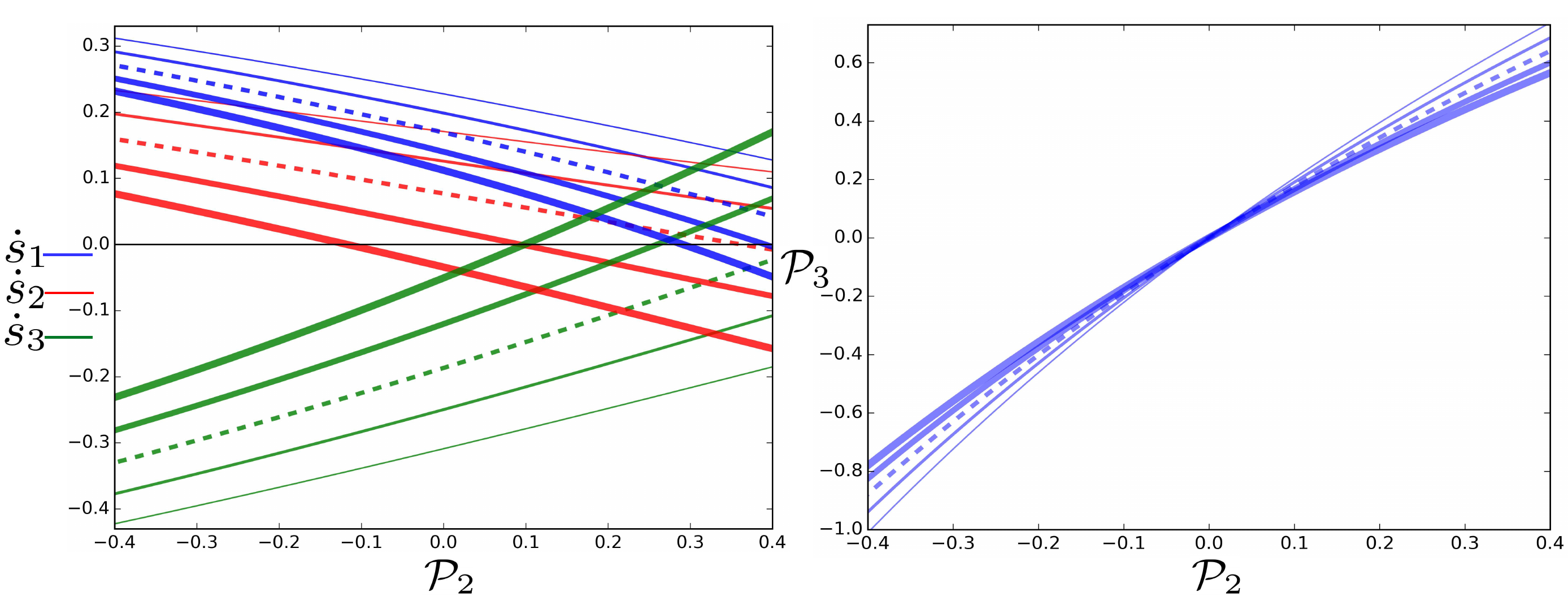}
\caption{The left panel shows how the growth/decrease rates $\dot{s}_i$ for strings depend on $\mathcal{P}_2$ values. The right panel shows the value of  $\mathcal{P}_3$ for the new (third) string. The thickness of lines shows variation of the value $\mathcal{P}_1$, i.e. the thinnest line corresponds to $\mathcal{P}_1=-0.4$ and the thickest one to $\mathcal{P}_1=0.4$ (dashed lines present the result for $\mathcal{P}_1=0.0$). Calculations are carried out for string tensions $\mu_1=1$, $\mu_2=1.2$, $\mu_3=1.4$. The vectors $\textbf{b}_i$ are orthogonal to each other.}
\label{fig:P3SSS}
\end{figure*}

Differentiating (\ref{Varfg}) we obtain
\begin{equation}
\begin{split}
\label{VertCondD}
& \, (1+\dot{s}_i) \textbf{a}_{i}^{\prime} - (1-\dot{s}_i) \textbf{b}_i^{\prime} = 2\dot{\mathbf{X}}(t), \\ 
& F^{\prime}_I (\dot{s}_I-1) = 2 \dot{\Phi}(t) = F^{\prime}_3 (\dot{s}_3+1),
\end{split}
\end{equation}
where $X^{\mu} = \left( t, \; \textbf{X}(t) \right)$ and $I=1,2$.

As it was done above, manipulating vectors $\textbf{a}_i$ and $\textbf{b}_i$, it is possible to obtain the following equations for $\dot{s}_i$

\begin{equation}
\label{ABeq}  
\textbf{a}_{k}^{\prime} (1+\dot{s}_k) = -\frac{2}{\mu} \sum_i (1-\dot{s}_i) \mu_i \textbf{b}_i^{\prime} + (1-\dot{s}_k) \textbf{b}_k^{\prime}.
\end{equation}

Squaring equations (\ref{ABeq}) and using the normalization conditions (\ref{CollidingNorm}), (\ref{JunctionSolNorm}) for $|\textbf{a}_i^{\prime}|$ and $|\textbf{b}_i^{\prime}|$, we obtain equations for $\dot{s}_i$ as functions of $F^{\prime}_i$ and the angles between $\textbf{b}_i^{\prime}$. Since we know the values of $F_I^{\prime}$, to solve equations for $\dot{s}_i$ we need to find $F_3^{\prime}$. By using the second line in (\ref{VertCondD}) one can obtain 
\begin{equation}
\label{F3}
F_3^{\prime} = \frac{F_I^{\prime} (\dot{s}_I-1)}{\dot{s}_3+1}.
\end{equation}

To define the string junction completely, we also need to find $\mathcal{D}_3$, which can be achieved by using the second line of relations (\ref{Sum1}). It provides the following equation
\begin{equation}
\label{D3}
\mathcal{D}_3 = \frac{1}{\mu_3} \sum\limits_{I=1}^2\mu_I \mathcal{D}_I.
\end{equation}

We can solve numerically equations (\ref{ABeq}) taking into account conditions (\ref{F3}) and (\ref{D3}) for different values of the current $F_{I}^{\prime}$. The result of these calculations for orthogonal vectors $\textbf{b}_i^{\prime}$ is shown in figure \ref{fig:S3F3Symm}, where the orthogonality was chosen just for explicit demonstration of the current`s influence on string growth/decrease rate.

\subsection{Collisions of non-identical strings} \label{Collisions of non-identical strings}

In the situation when colliding strings are not identical, i.e. they have different values of tensions $\mu_1 \neq \mu_2$ or/and currents $F_{1}^{\prime} \neq F_{2}^{\prime}$, it is seen from the second line of (\ref{VertCondD}) that the system has an additional restriction for $\dot{s}_{I}$. This leads to an overdetermined system of equations (\ref{Sum1}), (\ref{VertCondD}). A similar problem appeared in the work \cite{SteerLilleyYamauchiHiramatsu}, where the magnetic and electric regimes were studied. 

A possible resolution of this issue is to assume that vectors $\textbf{b}_i^{\prime}$ are not completely independent. We need to fix at least one angle between the vectors $\textbf{b}_i^{\prime}$. This means that this angle should have a specific value during the whole evolution, in contrast to angles for strings without currents. Such an example is illustrated in figure \ref{fig:S3F3NonSymm}.

We should notice that there is another possible way to escape the situation when equations for the junction are overdetermined. As it was described in \cite{RybakAvgoustidisMartins} the properties of strings, such as tension, mass per unit length and equations of motion depend on the product  $\mathcal{D} F^{\prime \, 2}$. It is always possible to multiply the function $F^{\prime \, 2}$ by a constant and divide $\mathcal{D}$ by the same constant, keeping all string properties and the equations of motion unchanged. In other words, imposing string properties we restrict the product $\mathcal{D} F^{\prime \, 2}$ rather than $F^{\prime \, 2}$ and $\mathcal{D}$ separately. This kind of rescaling provides an additional degree of freedom to avoid the situation when the junction equations (\ref{BoundTerms})-(\ref{Varfg}) are overdetermined. 
 
Let's define $\mathcal{P}_i \equiv \mathcal{D}_i |F_i^{\prime}|^2$. The values of colliding strings $\mathcal{P}_1$, $\mathcal{P}_2$ are known. We need to define $\mathcal{P}_3$ for a new string, which can be obtained from equations (\ref{Sum1}) and (\ref{VertCondD})

\begin{equation}
\label{P3} 
\mathcal{P}_3 = \frac{1}{\mu_3 (\dot{s}_3+1)^2} \sum\limits_{I=1}^{2} \mu_I \mathcal{P}_I (\dot{s}_I-1)^2.
\end{equation}

Equations (\ref{VertCondD}) can be automatically satisfied by a proper adjustment of $F^{\prime}_{i}$ and $\mathcal{D}_i$ that leaves the value of $\mathcal{P}_i$ unchanged. Substituting the equality (\ref{P3}) into equations (\ref{ABeq}) one can find $\dot{s}_i$ and $\mathcal{P}_3$. The result of such calculations is shown in figure \ref{fig:P3SSS}. There is no necessity to determine $F^{\prime}_{i}$ and $\mathcal{D}_i$ for the strings explicitly, since only their combinations in the form of $\mathcal{P}_i$ have an influence on the dynamics of junctions.

\section{Collision of straight chiral current-carrying strings} \label{Collision of straight strings with currents}

\subsection{Collisions of straight identical strings} \label{Collisions of straight identical strings}

Similarly as it was done above, we can consider a collision of straight strings with currents and find conditions which allow the formation of a junction in Minkowski space. Let's firstly consider the collision of two identical strings $\mu_1=\mu_2$, $\mathcal{D}_1=\mathcal{D}_2$, $F^{\prime}_1=F^{\prime}_2$.

In the first place we need to build an appropriate solution for straight strings with currents, i.e we need to find vectors $\textbf{x}_{1,2}$ for straight strings that can satisfy the previously described properties.
It can be done in the following way
\begin{equation}
\label{StrStringCurrent}
  \textbf{y}_{i} \equiv \textbf{x}_{i} + \textbf{z}_{i},
\end{equation}
where vectors $\textbf{x}_{i}$ are defined in the same way as for ordinary straight strings in Minkowski space \cite{CopelandKibbleSteer,CopelandKibbleSteer2}
\begin{equation}
\begin{split}
\label{StrStringCurrent1}
 & \textbf{x}_{1,2} = \left\lbrace -\gamma^{-1} \sigma \cos \alpha; \,\mp \gamma^{-1} \sigma \sin \alpha; \, \pm  \upsilon \tau \right\rbrace, \\
 & \qquad \qquad \qquad \textbf{x}_{3} = \left\lbrace \sigma; \, 0; \, 0 \right\rbrace.
\end{split}  
\end{equation}

Now we need to understand which form of vectors $\textbf{z}_i$ should be chosen in order to be in agreement with properties (\ref{VectorNorm}), (\ref{UsefulQuant}), (\ref{ConditionPhi}). Using the form of the string (\ref{StrStringCurrent1}), we can try the following ansatz
\begin{equation}
\begin{split}
\label{StrStringCurrent2}
 &  \textbf{z}_{1,2} =  f_{1,2}(\sigma-\tau) \big\lbrace \gamma^{-1}  \cos \alpha; \,\pm \gamma^{-1} \sin \alpha; \, \pm  \upsilon \tau \big\rbrace, \\
 & \qquad \qquad \qquad   \textbf{z}_{3} = \left\lbrace -f_3(\sigma+\tau); \, 0; \, 0 \right\rbrace,
\end{split}  
\end{equation}
where $f_i$ are arbitrary functions.

One can form ingoing and outgoing components from (\ref{StringSolution})
\begin{equation}
\begin{split}
\label{abCurrent1}
& \textbf{b}^{\prime}_i = \textbf{y}_i^{\prime} - \dot{\textbf{y}}_i,\\
& \textbf{a}^{\prime}_i = \textbf{y}_i^{\prime} + \dot{\textbf{y}}_i,
\end{split}
\end{equation}
which are normalized as
\begin{equation}
\begin{split}
\label{abCurrent2}
& \textbf{b}^{\prime \, 2}_{1,2} = 1 - 4 f_1^{\prime} (1-f_1^{\prime}), \quad \textbf{b}^{\prime \, 2}_3=1, \\
& \textbf{a}^{\prime \, 2}_{1,2} = 1, \quad \textbf{a}^{\prime \, 2}_{3} = 1 - 4 f_3^{\prime} (1-f_3^{\prime}).
\end{split}
\end{equation}

Comparing the result for straight strings with equalities (\ref{VectorNorm}), (\ref{UsefulQuant}) and (\ref{ConditionPhi}), we can conclude that 
\begin{equation}
\label{CurrentCon}
\mathcal{D}_i F_{i}^{\prime \, 2} = 4 f_i^{\prime}(1-f_i^{\prime}) =  \mathcal{P}_i.
\end{equation}

Let's consider the linearised contribution from the current
\begin{equation}
\begin{split}
\label{LinCurrent}
& f_{1,2} = \frac{\sqrt{1-\varphi_{1,2}}-1}{2} \;(\sigma - \tau), \\
& \; \; \; f_{3} = \frac{\sqrt{1-\varphi_{3}}-1}{2} \;(\sigma + \tau),
\end{split}
\end{equation}
where $\varphi_i$ are constants and for straight strings are determined as $\varphi_i = \mathcal{P}_i$.

\begin{figure}[ht]
\centering
\includegraphics[width=0.45\textwidth]{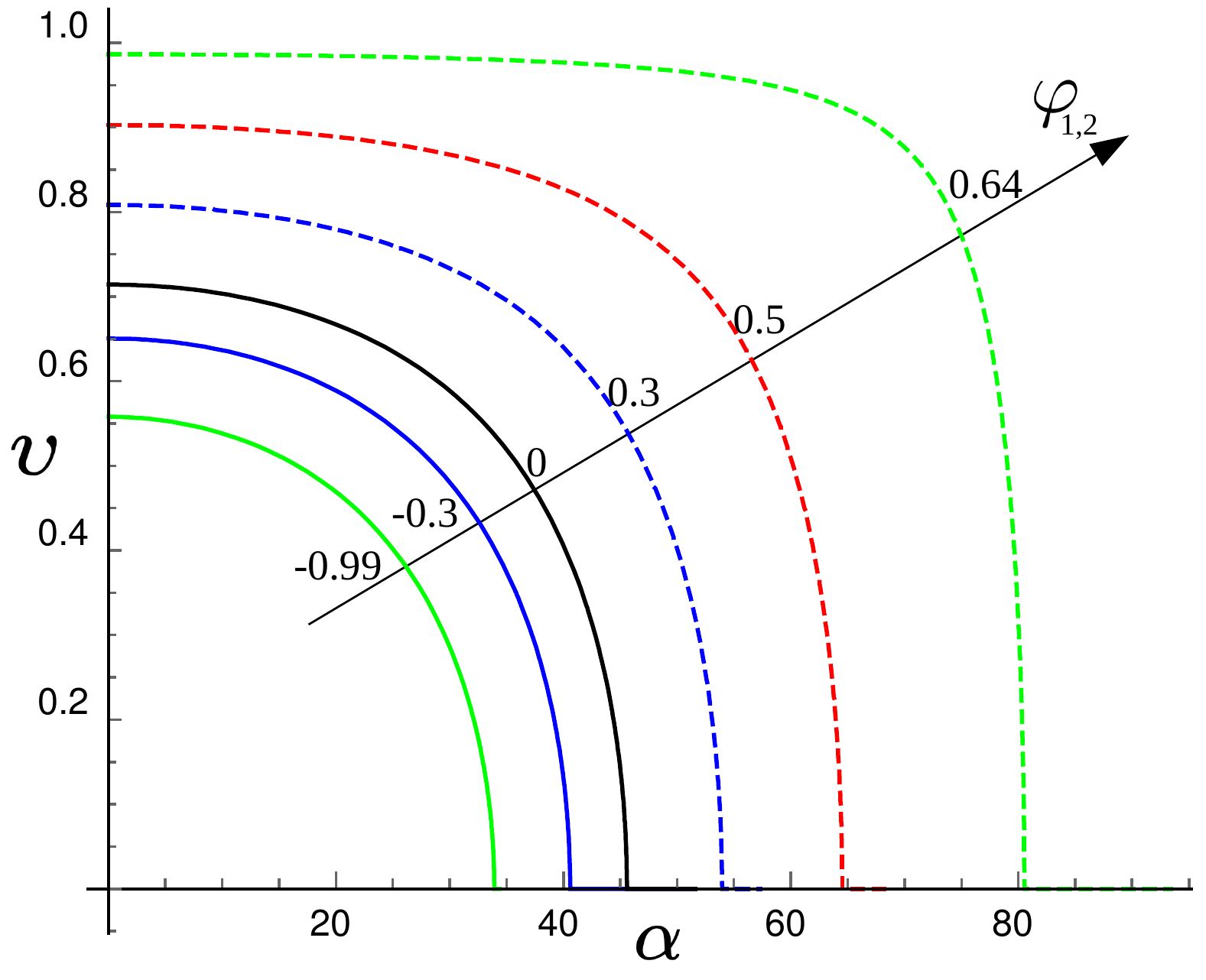}
\caption{\label{fig:areasCurrent} Range of parameters: velocity $\upsilon$ and angle $\alpha$, which allow the junction production ($\dot{s}_3>0$) for the case when the heaviest string has the tension $\mu_3=1.4 \, \mu_1=1.4 \, \mu_2$. It is seen how the region of junction production depends on the value of colliding strings currents (we also assumed $\varphi_{1}=\varphi_2$).}
\end{figure}

Applying the same method as in the work \cite{CopelandKibbleSteer2}, we can solve the kinematic equations (\ref{ABeq}) for straight strings with currents and find out for which range of velocities $\upsilon$ and angles $\alpha$ the production of a junction is possible ($\dot{s}_3>0$). Solutions for different values of the current are shown in figure \ref{fig:areasCurrent}.

As we see in figure \ref{fig:areasCurrent}, when the contribution from the current $\varphi_{1,2}$ increases the area that allows junctions production is growing. Decrease of the value $\varphi_{1,2}$ leads to smaller area of junctions production. While the negative value of the current can reach its minimum $\varphi_{1,2}=-1$, the positive value of the current is limited by the relation of string tensions. We are going to estimate this limit. Let's consider the string collision for the angle $\alpha=0$. For this particular situation, using equations (\ref{ABeq}), one can show that the physically meaningful solution for velocity is given by the expression

\begin{equation}
\begin{split}
\label{VelZeroAngl}
& \qquad \qquad \quad  v^2_{\alpha=0} = (2 \mu_{1} - \mu_3) \times \\
& \frac{\left( \varphi_{1} (2 \frac{\mu_{1}}{\mu_3} -1) + (\frac{\mu_{3}}{\mu_1} + 2) (1 + \sqrt{1-2 \varphi_{1} \frac{\mu_{1}}{\mu_3}} \right) }{8(1-\varphi_{1}) \mu_1},
\end{split}
\end{equation}
while the solution for $\cos \alpha$ when $v=0$ is given as
\begin{equation}
\begin{split}
\label{AnglZeroVel}
&  \cos \alpha_{v=0} = \frac{ \mu_3-2 \mu_1 + (2 \mu_1 + \mu_3) \sqrt{1 - 2 \varphi_1 \frac{\mu_1}{\mu_3}} }{4 \sqrt{1-\varphi_1} \mu_1}.
\end{split}
\end{equation}

From solutions (\ref{VelZeroAngl}) and (\ref{AnglZeroVel}) we can conclude that the contribution from the current is restricted by the equality
\begin{equation}
\label{MaximalVarphi}
\varphi_{\text{max}} = \frac{\mu_3}{2 \mu_1}.
\end{equation}

At the same time, we should notice that the maximal velocity is reached when
\begin{equation}
\label{MaximalVel}
\varphi_{v=1} = \frac{4 \mu_3^2}{(2 \mu_1 + \mu_3)^2},
\end{equation}
which is smaller than the maximal possible value of $\varphi_{\text{max}}$.

%%%%%%%%%%%%%%%%%%%%%%%%%%%%%%%%%%%%%%%%%%%%%%%%%%%%%%%%%%%%%%%%%%%%%%%%%%%%%%%%%%

\subsection{Collisions of straight non-identical strings} \label{Collisions of straight non-identical strings}

Here we use our freedom to change $F^{\prime}_i$ and $\mathcal{D}_i$ keeping the value $\mathcal{P}_i=\mathcal{D}_i |F^{\prime}_i|^2 $ constant. In this situation we avoid that equations for junctions are overdetermined and we can consider the collision of straight strings with different tensions. Let's start from the following ansatz 
\begin{equation}
\begin{split}
\label{StrStringCurrent3}
 & \textbf{x}_{1,2} = \left\lbrace -\gamma^{-1} \sigma \cos \alpha; \,\mp \gamma^{-1} \sigma \sin \alpha; \, \pm  \upsilon \tau \right\rbrace, \\
 & \quad \; \textbf{x}_{3} = \left\lbrace \gamma^{-1}_u \sigma \cos \theta; \, \gamma^{-1}_u \sigma \sin \theta; \, u \tau \right\rbrace,
\end{split}  
\end{equation}
where $\gamma^{-1}_u = \sqrt{1-u^2}$, $u$ is the third string velocity and the angle $\theta$ defines the orientation of this string.

The current-carrying ansatz we use has the form
\begin{equation}
\begin{split}
\label{StrStringCurrent4}
 & \textbf{z}_{1,2} = f_{1,2}(\sigma-\tau) \left\lbrace -\gamma^{-1} \cos \alpha; \,\mp \gamma^{-1} \sin \alpha; \, \pm  \upsilon \right\rbrace, \\
 & \quad \; \; \; \textbf{z}_{3} = f_{3}(\sigma+\tau) \left\lbrace \gamma^{-1}_u \cos \theta; \, \gamma^{-1}_u \sin \theta; \, u \right\rbrace.
\end{split}  
\end{equation}

We have two additional undefined parameters which characterize the new string: $u$ and $\theta$. To determine them we are going to use the equations for the junction dynamics. To do so we need to take the derivative of $\textbf{y}_3$ with respect to time at the point $\sigma_3 = s_3(\tau)$ 

\begin{equation}
\begin{split}
\label{Xjunct}
& \; \frac{d \textbf{y}_3(s_3(\tau), \tau)}{d \tau} \equiv \dot{\textbf{X}} = \left[ (\dot{s}_3 + 1) f_3^{\prime} + \dot{s}_3 \right] \times \\
& \left\lbrace \gamma^{-1}_u \cos \theta; \;  \gamma^{-1}_u \sin \theta; \; u \frac{(\dot{s}_3+1) f^{\prime}_3 +1}{(\dot{s}_3+1) f_3^{\prime} + \dot{s}_3}  \right\rbrace
\end{split}
\end{equation}
and use equations (\ref{VertCondD}), (\ref{ABeq}) to express the  vector $\dot{\textbf{X}}$ as a linear combination of vectors $\textbf{b}^{\prime}_i$. These additional equations define the angle $\theta$ and velocity $u$ for the third string. 

\begin{figure}[ht]
\centering
\includegraphics[width=0.46\textwidth]{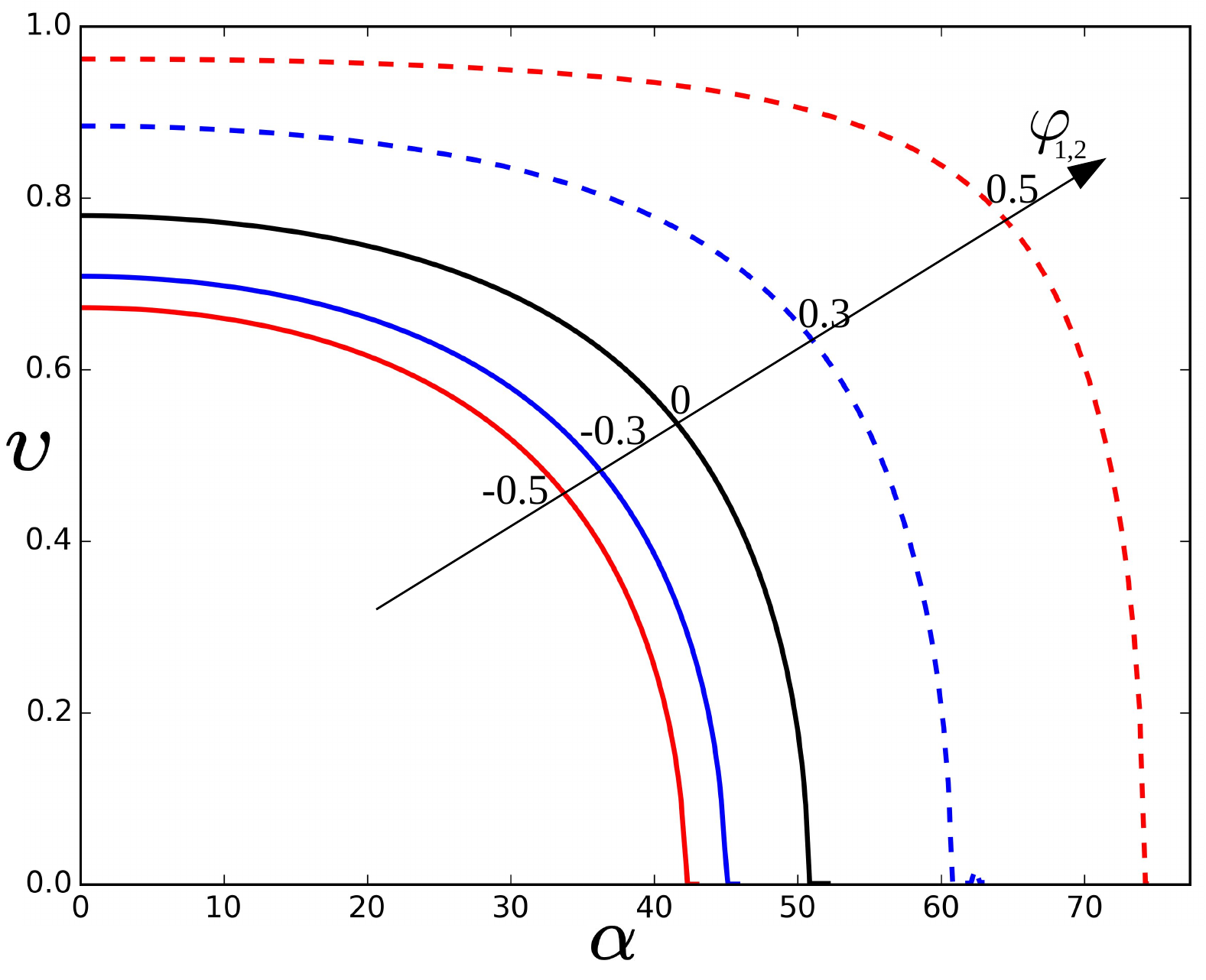}
\caption{\label{fig:areasCurrentAssym} Range of parameters: velocity $\upsilon$ and angle $\alpha$, which allow the junction production ($\dot{s}_3>0$) for the case when string tensions are $\mu_1=1$,  $\mu_2=1.2$,  $\mu_3 = 1.4$. It is seen how the region of junction production depends on the value of colliding strings currents (we also assumed $\varphi_{1}=\varphi_2$).}
\end{figure}

Now we can repeat the same procedure as in the section \ref{Collisions of straight identical strings}: we build vectors $\textbf{y}_i$ by using (\ref{StrStringCurrent}), calculate ingoing and outgoing parts from (\ref{abCurrent1}) and obtain that conditions (\ref{abCurrent2}) are valid for vectors (\ref{StrStringCurrent3}), (\ref{StrStringCurrent4}) as well. We use the linear contribution from the current in the form (\ref{LinCurrent}). Defining the values of the third string $\mathcal{P}_3$ by (\ref{CurrentCon}) we use equations (\ref{ABeq}), (\ref{P3}) together with (\ref{VertCondD}), (\ref{ABeq}) and (\ref{Xjunct}) complete the calculation and find out for which $v$ and $\alpha$ the growth rate $\dot{s}_3=0$. The result is shown in figure \ref{fig:areasCurrentAssym}.

\section{Conclusions} \label{Conclusions}

In section \ref{Solution in Minkowski space for strings with chiral currents} we extended the exact solution for chiral superconducting strings in Minkowski space for the generalized string action of the form (\ref{Action}). The final solution is different from previously studied particular cases \cite{CarterPeter2, DavisKibblePicklesSteer, Blanco-PilladoOlumVilenkin} only by the constant multiplier $\mathcal{D}$. This constant is defined by the form of the current dependence in the the action.

Using the solution from section \ref{Solution in Minkowski space for strings with chiral currents}, we considered junction dynamics for strings with chiral currents. We used the assumption that the current on the third string (the string created by the collision) is not defined by fundamental principles and can be treated as a free parameter that can be found from kinematic constraints. We showed that when two identical strings collide it is possible to describe the dynamics of a junction without having an overdetermined system of equations. At the same time, if two colliding strings are not identical, the equations impose additional constraints for the ingoing vectors of string solutions $\textbf{b}^{\prime}_i$.

Meanwhile we should remember that the string action (\ref{Action}) is just an effective way to describe strings with currents. This fact implies that if we want to guarantee given string properties we do not need to define both string parameters $F^{\prime}$ and $\mathcal{D}$. We have a rescaling freedom to make $F^{\prime}$ and $\mathcal{D}$ arbitrary, keeping the combination $F^{\prime \, 2} \mathcal{D}$ unchanged. This transformation freedom allowed us to consider junction dynamics without establishing any correlations for vectors $\textbf{b}^{\prime}_i$.

It is important to specify the difference between chiral strings, considered in this  paper, and electric/magnetic strings studied in \cite{SteerLilleyYamauchiHiramatsu}. Due to the chiral conditions (Eq. \ref{ChiralCurrent}), the different Lagrangians (with different values of the coupling constants or functional forms of the $\kappa$ and $\Delta$ dependence) contribute to the analytic solution of the equations of motion and equation of state only as additional multipliers, which is not the case for electric and magnetic strings. If these additional multipliers $\mathcal{D}$ and values of currents $F^{\prime \, 2}$  are not fixed separately (since they appear in equation of state only as a combination $F^{\prime \, 2} \mathcal{D}$), they allow enough freedom to avoid the overdetermination issue for the case of the chiral current. This approach is specific to the chiral type of current and won't change the previously obtained result for magnetic and electric cases \cite{SteerLilleyYamauchiHiramatsu}.

Despite of different possible resolutions to avoid overdetermined equations, we found out that in all cases the influence of currents has a similar tendency: the rate of strings growth/decrease $|\dot{s}_i|$ slows down when the string current increases. Moreover, some values of the current can even swap the junction dynamics, i.e. heavier strings for particular configurations can start to grow instead of decreasing. This feature might play an important role for the string network evolution. 
 
In section \ref{Junctions with chiral currents} we studied the collision of straight superconducting chiral strings. We demonstrated that an increase of the current makes the area of junctions production bigger, while a decrease leads to the smaller ``velocity-angle" region where strings can create a Y junction. Additionally, we noticed that there is a maximal value for the current contribution that still can allow production of junctions, while the negative value is not limited and can reach up to $\varphi=-1$.

To conclude, we want to highlight that in spite of the possible resolution for the overdetermined system of equations, this problem needs further investigation. It should be clarified whether it is possible to implement additional constraints on the junction properties or maybe in contrast, whether it is possible to relax the imposed conditions by new degrees of freedom. Another approach should come from numerical simulations in field theory \cite{MoriartyMyersRebbi, LagunaRichard}. New computations in this direction would be particularly useful, shedding more light on this intriguing problem.

\begin{acknowledgments}

This work was financed by FEDER---Fundo Europeu de Desenvolvimento Regional funds through the COMPETE 2020---Operational Programme for Competitiveness and Internationalisation (POCI), and by Portuguese funds through FCT---Funda\c c\~ao para a Ci\^encia e a Tecnologia in the framework of the project POCI-01-0145-FEDER-028987. The work of IR is supported by the FCT fellowship SFRH/BD/52699/2014, within the FCT Doctoral Program PhD::SPACE (PD/00040/2012). The work of AA was partly supported by an Advanced Nottingham Research Fellowship and an STFC consolidator grant at the University of Nottingham.

\end{acknowledgments}

\bibliography{junctions}

\begin{thebibliography}{48}%
\makeatletter
\providecommand \@ifxundefined [1]{%
 \@ifx{#1\undefined}
}%
\providecommand \@ifnum [1]{%
 \ifnum #1\expandafter \@firstoftwo
 \else \expandafter \@secondoftwo
 \fi
}%
\providecommand \@ifx [1]{%
 \ifx #1\expandafter \@firstoftwo
 \else \expandafter \@secondoftwo
 \fi
}%
\providecommand \natexlab [1]{#1}%
\providecommand \enquote  [1]{``#1''}%
\providecommand \bibnamefont  [1]{#1}%
\providecommand \bibfnamefont [1]{#1}%
\providecommand \citenamefont [1]{#1}%
\providecommand \href@noop [0]{\@secondoftwo}%
\providecommand \href [0]{\begingroup \@sanitize@url \@href}%
\providecommand \@href[1]{\@@startlink{#1}\@@href}%
\providecommand \@@href[1]{\endgroup#1\@@endlink}%
\providecommand \@sanitize@url [0]{\catcode `\\12\catcode `\$12\catcode
  `\&12\catcode `\#12\catcode `\^12\catcode `\_12\catcode `\%12\relax}%
\providecommand \@@startlink[1]{}%
\providecommand \@@endlink[0]{}%
\providecommand \url  [0]{\begingroup\@sanitize@url \@url }%
\providecommand \@url [1]{\endgroup\@href {#1}{\urlprefix }}%
\providecommand \urlprefix  [0]{URL }%
\providecommand \Eprint [0]{\href }%
\providecommand \doibase [0]{http://dx.doi.org/}%
\providecommand \selectlanguage [0]{\@gobble}%
\providecommand \bibinfo  [0]{\@secondoftwo}%
\providecommand \bibfield  [0]{\@secondoftwo}%
\providecommand \translation [1]{[#1]}%
\providecommand \BibitemOpen [0]{}%
\providecommand \bibitemStop [0]{}%
\providecommand \bibitemNoStop [0]{.\EOS\space}%
\providecommand \EOS [0]{\spacefactor3000\relax}%
\providecommand \BibitemShut  [1]{\csname bibitem#1\endcsname}%
\let\auto@bib@innerbib\@empty
%</preamble>
\bibitem [{\citenamefont {Kibble}(1976)}]{Kibble}%
  \BibitemOpen
  \bibfield  {author} {\bibinfo {author} {\bibfnamefont {T.~W.~B.}\
  \bibnamefont {Kibble}},\ }\href {http://stacks.iop.org/0305-4470/9/i=8/a=029}
  {\bibfield  {journal} {\bibinfo  {journal} {Journal of Physics A:
  Mathematical and General}\ }\textbf {\bibinfo {volume} {9}},\ \bibinfo
  {pages} {1387} (\bibinfo {year} {1976})}\BibitemShut {NoStop}%
\bibitem [{\citenamefont {Burgess}\ \emph {et~al.}(2001)\citenamefont
  {Burgess}, \citenamefont {Majumdar}, \citenamefont {Nolte}, \citenamefont
  {Quevedo}, \citenamefont {Rajesh},\ and\ \citenamefont
  {Zhang}}]{BurgessMajumdarNolteQuevedoRajeshZhang}%
  \BibitemOpen
  \bibfield  {author} {\bibinfo {author} {\bibfnamefont {C.~P.}\ \bibnamefont
  {Burgess}}, \bibinfo {author} {\bibfnamefont {M.}~\bibnamefont {Majumdar}},
  \bibinfo {author} {\bibfnamefont {D.}~\bibnamefont {Nolte}}, \bibinfo
  {author} {\bibfnamefont {F.}~\bibnamefont {Quevedo}}, \bibinfo {author}
  {\bibfnamefont {G.}~\bibnamefont {Rajesh}}, \ and\ \bibinfo {author}
  {\bibfnamefont {R.-J.}\ \bibnamefont {Zhang}},\ }\href {\doibase
  10.1088/1126-6708/2001/07/047} {\bibfield  {journal} {\bibinfo  {journal}
  {JHEP}\ }\textbf {\bibinfo {volume} {07}},\ \bibinfo {pages} {047} (\bibinfo
  {year} {2001})},\ \Eprint {http://arxiv.org/abs/hep-th/0105204}
  {arXiv:hep-th/0105204 [hep-th]} \BibitemShut {NoStop}%
%%CITATION = HEP-TH/0105204;%%
\bibitem [{\citenamefont {Dvali}\ \emph {et~al.}(2004)\citenamefont {Dvali},
  \citenamefont {Kallosh},\ and\ \citenamefont
  {Van~Proeyen}}]{DvaliKalloshProeyen}%
  \BibitemOpen
  \bibfield  {author} {\bibinfo {author} {\bibfnamefont {G.}~\bibnamefont
  {Dvali}}, \bibinfo {author} {\bibfnamefont {R.}~\bibnamefont {Kallosh}}, \
  and\ \bibinfo {author} {\bibfnamefont {A.}~\bibnamefont {Van~Proeyen}},\
  }\href {http://stacks.iop.org/1126-6708/2004/i=01/a=035} {\bibfield
  {journal} {\bibinfo  {journal} {JHEP}\ }\textbf {\bibinfo {volume} {2004}},\
  \bibinfo {pages} {035} (\bibinfo {year} {2004})},\ \Eprint
  {http://arxiv.org/abs/hep-th/0312005v3} {arXiv:hep-th/0312005v3 [hep-th]}
  \BibitemShut {NoStop}%
%%CITATION = hep-th/0312005v3;%%
\bibitem [{\citenamefont {Dvali}\ and\ \citenamefont
  {Vilenkin}(2004)}]{DvaliVilenkin}%
  \BibitemOpen
  \bibfield  {author} {\bibinfo {author} {\bibfnamefont {G.}~\bibnamefont
  {Dvali}}\ and\ \bibinfo {author} {\bibfnamefont {A.}~\bibnamefont
  {Vilenkin}},\ }\href {http://stacks.iop.org/1475-7516/2004/i=03/a=010}
  {\bibfield  {journal} {\bibinfo  {journal} {JCAP}\ }\textbf {\bibinfo
  {volume} {2004}},\ \bibinfo {pages} {010} (\bibinfo {year} {2004})},\ \Eprint
  {http://arxiv.org/abs/hep-th/0312007v2} {arXiv:hep-th/0312007v2 [hep-th]}
  \BibitemShut {NoStop}%
%%CITATION = hep-th/0312007v2;%%
\bibitem [{\citenamefont {Copeland}\ \emph {et~al.}(2004)\citenamefont
  {Copeland}, \citenamefont {Myers},\ and\ \citenamefont
  {Polchinski}}]{PolchinskiCopelandMyers}%
  \BibitemOpen
  \bibfield  {author} {\bibinfo {author} {\bibfnamefont {E.~J.}\ \bibnamefont
  {Copeland}}, \bibinfo {author} {\bibfnamefont {R.~C.}\ \bibnamefont {Myers}},
  \ and\ \bibinfo {author} {\bibfnamefont {J.}~\bibnamefont {Polchinski}},\
  }\href {\doibase 10.1088/1126-6708/2004/06/013} {\bibfield  {journal}
  {\bibinfo  {journal} {JHEP}\ }\textbf {\bibinfo {volume} {2004}},\ \bibinfo
  {pages} {013} (\bibinfo {year} {2004})},\ \Eprint
  {http://arxiv.org/abs/hep-th/0312067v5} {arXiv:hep-th/0312067v5 [hep-th]}
  \BibitemShut {NoStop}%
%%CITATION = ARXIV: hep-th/0312067v5;%%
\bibitem [{\citenamefont {Sarangi}\ and\ \citenamefont
  {Tye}(2002)}]{SarangiTye}%
  \BibitemOpen
  \bibfield  {author} {\bibinfo {author} {\bibfnamefont {S.}~\bibnamefont
  {Sarangi}}\ and\ \bibinfo {author} {\bibfnamefont {S.-H.~H.}\ \bibnamefont
  {Tye}},\ }\href {\doibase 10.1016/S0370-2693(02)01824-5} {\bibfield
  {journal} {\bibinfo  {journal} {Phys.Lett.B}\ }\textbf {\bibinfo {volume}
  {536}},\ \bibinfo {pages} {185} (\bibinfo {year} {2002})},\ \Eprint
  {http://arxiv.org/abs/hep-th/0204074v1} {arXiv:hep-th/0204074v1 [hep-th]}
  \BibitemShut {NoStop}%
%%CITATION = ARXIV: hep-th/0204074v1;%%
\bibitem [{\citenamefont {Firouzjahi}\ and\ \citenamefont
  {Tye}(2005)}]{FirouzjahiTye}%
  \BibitemOpen
  \bibfield  {author} {\bibinfo {author} {\bibfnamefont {H.}~\bibnamefont
  {Firouzjahi}}\ and\ \bibinfo {author} {\bibfnamefont {S.-H.~H.}\ \bibnamefont
  {Tye}},\ }\href {\doibase 10.1088/1475-7516/2005/03/009} {\bibfield
  {journal} {\bibinfo  {journal} {JCAP}\ }\textbf {\bibinfo {volume} {0503}},\
  \bibinfo {pages} {009} (\bibinfo {year} {2005})},\ \Eprint
  {http://arxiv.org/abs/hep-th/0501099v3} {arXiv:hep-th/0501099v3 [hep-th]}
  \BibitemShut {NoStop}%
%%CITATION = hep-th/0501099v3;%%
\bibitem [{\citenamefont {Jones}\ \emph {et~al.}(2003)\citenamefont {Jones},
  \citenamefont {Stoica},\ and\ \citenamefont {Tye}}]{JonesStoicaTye}%
  \BibitemOpen
  \bibfield  {author} {\bibinfo {author} {\bibfnamefont {N.~T.}\ \bibnamefont
  {Jones}}, \bibinfo {author} {\bibfnamefont {H.}~\bibnamefont {Stoica}}, \
  and\ \bibinfo {author} {\bibfnamefont {S.-H.~H.}\ \bibnamefont {Tye}},\
  }\href {\doibase 10.1016/S0370-2693(03)00592-6} {\bibfield  {journal}
  {\bibinfo  {journal} {Phys.Lett.}\ }\textbf {\bibinfo {volume} {B}},\
  \bibinfo {pages} {6} (\bibinfo {year} {2003})},\ \Eprint
  {http://arxiv.org/abs/hep-th/0303269v1} {arXiv:hep-th/0303269v1 [hep-th]}
  \BibitemShut {NoStop}%
%%CITATION = hep-th/0303269v1;%%
\bibitem [{\citenamefont {Jeannerot}\ \emph {et~al.}(2003)\citenamefont
  {Jeannerot}, \citenamefont {Rocher},\ and\ \citenamefont
  {Sakellariadou}}]{JeannerotRocherSakellariadou}%
  \BibitemOpen
  \bibfield  {author} {\bibinfo {author} {\bibfnamefont {R.}~\bibnamefont
  {Jeannerot}}, \bibinfo {author} {\bibfnamefont {J.}~\bibnamefont {Rocher}}, \
  and\ \bibinfo {author} {\bibfnamefont {M.}~\bibnamefont {Sakellariadou}},\
  }\href {\doibase 10.1103/PhysRevD.68.103514} {\bibfield  {journal} {\bibinfo
  {journal} {Phys.Rev.D}\ }\textbf {\bibinfo {volume} {68}},\ \bibinfo {pages}
  {103514} (\bibinfo {year} {2003})},\ \Eprint
  {http://arxiv.org/abs/hep-ph/0308134v1} {arXiv:hep-ph/0308134v1 [hep-ph]}
  \BibitemShut {NoStop}%
%%CITATION = ARXIV: hep-ph/0308134v1;%%
\bibitem [{\citenamefont {Cui}\ \emph {et~al.}(2008)\citenamefont {Cui},
  \citenamefont {Martin}, \citenamefont {Morrissey},\ and\ \citenamefont
  {Wells}}]{CuiMartinMorrisseyWells}%
  \BibitemOpen
  \bibfield  {author} {\bibinfo {author} {\bibfnamefont {Y.}~\bibnamefont
  {Cui}}, \bibinfo {author} {\bibfnamefont {S.}~\bibnamefont {Martin}},
  \bibinfo {author} {\bibfnamefont {D.~E.}\ \bibnamefont {Morrissey}}, \ and\
  \bibinfo {author} {\bibfnamefont {J.}~\bibnamefont {Wells}},\ }\href
  {\doibase 10.1103/PhysRevD.77.043528} {\bibfield  {journal} {\bibinfo
  {journal} {Phys.Rev.D}\ }\textbf {\bibinfo {volume} {77}},\ \bibinfo {pages}
  {043528} (\bibinfo {year} {2008})},\ \Eprint
  {http://arxiv.org/abs/0709.0950v2} {arXiv:0709.0950v2 [hep-ph]} \BibitemShut
  {NoStop}%
%%CITATION = ARXIV: arXiv:0709.0950v2;%%
\bibitem [{\citenamefont {Jeannerot}\ and\ \citenamefont
  {Postma}(2004)}]{JeannerotPostma}%
  \BibitemOpen
  \bibfield  {author} {\bibinfo {author} {\bibfnamefont {R.}~\bibnamefont
  {Jeannerot}}\ and\ \bibinfo {author} {\bibfnamefont {M.}~\bibnamefont
  {Postma}},\ }\href {\doibase 10.1088/1126-6708/2004/12/043} {\bibfield
  {journal} {\bibinfo  {journal} {JHEP}\ }\textbf {\bibinfo {volume} {0412}},\
  \bibinfo {pages} {043} (\bibinfo {year} {2004})},\ \Eprint
  {http://arxiv.org/abs/hep-ph/0411260} {arXiv:hep-ph/0411260 [astro-ph.CO]}
  \BibitemShut {NoStop}%
%%CITATION = hep-ph/0411260;%%
\bibitem [{\citenamefont {Ach\'ucarro}\ \emph {et~al.}(2006)\citenamefont
  {Ach\'ucarro}, \citenamefont {Celi}, \citenamefont {Esole}, \citenamefont
  {Van~den Bergh},\ and\ \citenamefont {A.}}]{AchucarroCeli}%
  \BibitemOpen
  \bibfield  {author} {\bibinfo {author} {\bibfnamefont {A.}~\bibnamefont
  {Ach\'ucarro}}, \bibinfo {author} {\bibfnamefont {A.}~\bibnamefont {Celi}},
  \bibinfo {author} {\bibfnamefont {M.}~\bibnamefont {Esole}}, \bibinfo
  {author} {\bibfnamefont {J.}~\bibnamefont {Van~den Bergh}}, \ and\ \bibinfo
  {author} {\bibfnamefont {V.~P.}\ \bibnamefont {A.}},\ }\href {\doibase
  10.1088/1126-6708/2006/01/102} {\bibfield  {journal} {\bibinfo  {journal}
  {JHEP}\ }\textbf {\bibinfo {volume} {0601}},\ \bibinfo {pages} {102}
  (\bibinfo {year} {2006})},\ \Eprint {http://arxiv.org/abs/hep-th/0511001v2}
  {arXiv:hep-th/0511001v2 [hep-th]} \BibitemShut {NoStop}%
%%CITATION = hep-th/0511001v2;%%
\bibitem [{\citenamefont {Majumdar}\ and\ \citenamefont
  {Davis}(2002)}]{DavisMajumdar}%
  \BibitemOpen
  \bibfield  {author} {\bibinfo {author} {\bibfnamefont {M.}~\bibnamefont
  {Majumdar}}\ and\ \bibinfo {author} {\bibfnamefont {A.~C.}\ \bibnamefont
  {Davis}},\ }\href {http://stacks.iop.org/1126-6708/2002/i=03/a=056}
  {\bibfield  {journal} {\bibinfo  {journal} {JHEP}\ }\textbf {\bibinfo
  {volume} {2002}},\ \bibinfo {pages} {056} (\bibinfo {year} {2002})},\ \Eprint
  {http://arxiv.org/abs/hep-th/0202148v3} {arXiv:hep-th/0202148v3 [hep-th]}
  \BibitemShut {NoStop}%
%%CITATION = hep-th/0202148v3;%%
\bibitem [{\citenamefont {Allys}(2016)}]{Allys}%
  \BibitemOpen
  \bibfield  {author} {\bibinfo {author} {\bibfnamefont {E.}~\bibnamefont
  {Allys}},\ }\href {\doibase 10.1088/1475-7516/2016/04/009} {\bibfield
  {journal} {\bibinfo  {journal} {JCAP}\ }\textbf {\bibinfo {volume} {1604}},\
  \bibinfo {pages} {009} (\bibinfo {year} {2016})},\ \Eprint
  {http://arxiv.org/abs/1505.07888v3} {arXiv:1505.07888v3 [gr-qc]} \BibitemShut
  {NoStop}%
%%CITATION = arXiv:1505.07888;%%
\bibitem [{\citenamefont {Koehn}\ and\ \citenamefont
  {Trodden}(2016)}]{KoehnTrodden}%
  \BibitemOpen
  \bibfield  {author} {\bibinfo {author} {\bibfnamefont {M.}~\bibnamefont
  {Koehn}}\ and\ \bibinfo {author} {\bibfnamefont {M.}~\bibnamefont
  {Trodden}},\ }\href {\doibase 10.1016/j.physletb.2016.02.067} {\bibfield
  {journal} {\bibinfo  {journal} {Phys.Lett.}\ }\textbf {\bibinfo {volume}
  {B}},\ \bibinfo {pages} {498} (\bibinfo {year} {2016})},\ \Eprint
  {http://arxiv.org/abs/1512.09138v1} {arXiv:1512.09138v1 [hep-th]}
  \BibitemShut {NoStop}%
%%CITATION = arXiv:1512.09138;%%
\bibitem [{\citenamefont {Lazarides}\ \emph {et~al.}(2008)\citenamefont
  {Lazarides}, \citenamefont {Peddie},\ and\ \citenamefont
  {Vamvasakis}}]{LazaridesPeddieVamvasakis}%
  \BibitemOpen
  \bibfield  {author} {\bibinfo {author} {\bibfnamefont {G.}~\bibnamefont
  {Lazarides}}, \bibinfo {author} {\bibfnamefont {I.~N.~R.}\ \bibnamefont
  {Peddie}}, \ and\ \bibinfo {author} {\bibfnamefont {A.}~\bibnamefont
  {Vamvasakis}},\ }\href {\doibase 10.1103/PhysRevD.78.043518} {\bibfield
  {journal} {\bibinfo  {journal} {Phys.Rev.}\ }\textbf {\bibinfo {volume}
  {D}},\ \bibinfo {pages} {043518} (\bibinfo {year} {2008})},\ \Eprint
  {http://arxiv.org/abs/0804.3661v2} {arXiv:0804.3661v2 [hep-ph]} \BibitemShut
  {NoStop}%
%%CITATION = arXiv:0804.3661v2;%%
\bibitem [{\citenamefont {Chernoff}\ and\ \citenamefont
  {Tye}(2015)}]{ChernoffTye}%
  \BibitemOpen
  \bibfield  {author} {\bibinfo {author} {\bibfnamefont {D.~F.}\ \bibnamefont
  {Chernoff}}\ and\ \bibinfo {author} {\bibfnamefont {S.-H.~H.}\ \bibnamefont
  {Tye}},\ }\href {\doibase 10.1142/S0218271815300104} {\bibfield  {journal}
  {\bibinfo  {journal} {Int.J.Mod.Phys.}\ }\textbf {\bibinfo {volume} {D24}},\
  \bibinfo {pages} {1530010} (\bibinfo {year} {2015})},\ \Eprint
  {http://arxiv.org/abs/1412.0579v2} {arXiv:1412.0579v2 [astro-ph.CO]}
  \BibitemShut {NoStop}%
%%CITATION = arXiv:1412.0579v2;%%
\bibitem [{\citenamefont {Brandenberger}(2014)}]{BRANDENBERGER}%
  \BibitemOpen
  \bibfield  {author} {\bibinfo {author} {\bibfnamefont {R.~H.}\ \bibnamefont
  {Brandenberger}},\ }\href {\doibase
  https://doi.org/10.1016/j.nuclphysbps.2013.10.064} {\bibfield  {journal}
  {\bibinfo  {journal} {Nuclear Physics B - Proceedings Supplements}\ }\textbf
  {\bibinfo {volume} {246-247}},\ \bibinfo {pages} {45 } (\bibinfo {year}
  {2014})},\ \bibinfo {note} {proceedings of the 9th International Symposium on
  Cosmology and Particle Astrophysics},\ \Eprint
  {http://arxiv.org/abs/1301.2856v1} {arXiv:1301.2856v1 [astro-ph.CO]}
  \BibitemShut {NoStop}%
%%CITATION = arXiv:1301.2856v1;%%
\bibitem [{\citenamefont {Brandenberger}(2013)}]{BRANDENBERGER2}%
  \BibitemOpen
  \bibfield  {author} {\bibinfo {author} {\bibfnamefont {R.~H.}\ \bibnamefont
  {Brandenberger}},\ }\href@noop {} {\bibfield  {journal} {\bibinfo  {journal}
  {The Universe}\ }\textbf {\bibinfo {volume} {1}},\ \bibinfo {pages} {6 }
  (\bibinfo {year} {2013})},\ \Eprint {http://arxiv.org/abs/1401.4619v1}
  {arXiv:1401.4619v1 [astro-ph.CO]} \BibitemShut {NoStop}%
%%CITATION = arXiv:1401.4619v1;%%
\bibitem [{\citenamefont {Kawasaki}\ \emph {et~al.}(2015)\citenamefont
  {Kawasaki}, \citenamefont {Saikawa},\ and\ \citenamefont
  {Sekiguchi}}]{KawasakiKen'ichiSekiguchi}%
  \BibitemOpen
  \bibfield  {author} {\bibinfo {author} {\bibfnamefont {M.}~\bibnamefont
  {Kawasaki}}, \bibinfo {author} {\bibfnamefont {K.}~\bibnamefont {Saikawa}}, \
  and\ \bibinfo {author} {\bibfnamefont {T.}~\bibnamefont {Sekiguchi}},\ }\href
  {\doibase 10.1103/PhysRevD.91.065014} {\bibfield  {journal} {\bibinfo
  {journal} {Phys. Rev. D}\ }\textbf {\bibinfo {volume} {91}},\ \bibinfo
  {pages} {065014} (\bibinfo {year} {2015})},\ \Eprint
  {http://arxiv.org/abs/1412.0789v3} {arXiv:1412.0789v3 [hep-ph]} \BibitemShut
  {NoStop}%
%%CITATION = arXiv:1412.0789v3;%%
\bibitem [{\citenamefont {Gripaios}\ and\ \citenamefont
  {Randal-Williams}(2018)}]{GripaiosRandal-Williams}%
  \BibitemOpen
  \bibfield  {author} {\bibinfo {author} {\bibfnamefont {B.}~\bibnamefont
  {Gripaios}}\ and\ \bibinfo {author} {\bibfnamefont {O.}~\bibnamefont
  {Randal-Williams}},\ }\href {\doibase
  https://doi.org/10.1016/j.physletb.2018.05.013} {\bibfield  {journal}
  {\bibinfo  {journal} {Physics Letters B}\ }\textbf {\bibinfo {volume}
  {782}},\ \bibinfo {pages} {94 } (\bibinfo {year} {2018})},\ \Eprint
  {http://arxiv.org/abs/1610.05623v2} {arXiv:1610.05623v2 [hep-th]}
  \BibitemShut {NoStop}%
%%CITATION = arXiv:1610.05623v2;%%
\bibitem [{\citenamefont {Copeland}\ \emph {et~al.}(2006)\citenamefont
  {Copeland}, \citenamefont {Kibble},\ and\ \citenamefont
  {Steer}}]{CopelandKibbleSteer}%
  \BibitemOpen
  \bibfield  {author} {\bibinfo {author} {\bibfnamefont {E.~J.}\ \bibnamefont
  {Copeland}}, \bibinfo {author} {\bibfnamefont {T.~W.~B.}\ \bibnamefont
  {Kibble}}, \ and\ \bibinfo {author} {\bibfnamefont {D.~A.}\ \bibnamefont
  {Steer}},\ }\href {\doibase 10.1103/PhysRevLett.97.021602} {\bibfield
  {journal} {\bibinfo  {journal} {Phys. Rev. Lett.}\ }\textbf {\bibinfo
  {volume} {97}},\ \bibinfo {pages} {021602} (\bibinfo {year} {2006})},\
  \Eprint {http://arxiv.org/abs/hep-th/0601153v3} {arXiv:hep-th/0601153v3
  [hep-th]} \BibitemShut {NoStop}%
%%CITATION = hep-th/0601153v3;%%
\bibitem [{\citenamefont {Copeland}\ \emph {et~al.}(2007)\citenamefont
  {Copeland}, \citenamefont {Kibble},\ and\ \citenamefont
  {Steer}}]{CopelandKibbleSteer2}%
  \BibitemOpen
  \bibfield  {author} {\bibinfo {author} {\bibfnamefont {E.~J.}\ \bibnamefont
  {Copeland}}, \bibinfo {author} {\bibfnamefont {T.~W.~B.}\ \bibnamefont
  {Kibble}}, \ and\ \bibinfo {author} {\bibfnamefont {D.~A.}\ \bibnamefont
  {Steer}},\ }\href {\doibase 10.1103/PhysRevD.75.065024} {\bibfield  {journal}
  {\bibinfo  {journal} {Phys. Rev. D}\ }\textbf {\bibinfo {volume} {75}},\
  \bibinfo {pages} {065024} (\bibinfo {year} {2007})},\ \Eprint
  {http://arxiv.org/abs/hep-th/0611243v2} {arXiv:hep-th/0611243v2 [hep-th]}
  \BibitemShut {NoStop}%
%%CITATION = hep-th/0611243v2;%%
\bibitem [{\citenamefont {Copeland}\ \emph {et~al.}(2008)\citenamefont
  {Copeland}, \citenamefont {Firouzjahi}, \citenamefont {Kibble},\ and\
  \citenamefont {Steer}}]{CopelandFirouzjahiKibbleSteer}%
  \BibitemOpen
  \bibfield  {author} {\bibinfo {author} {\bibfnamefont {E.~J.}\ \bibnamefont
  {Copeland}}, \bibinfo {author} {\bibfnamefont {H.}~\bibnamefont
  {Firouzjahi}}, \bibinfo {author} {\bibfnamefont {T.~W.~B.}\ \bibnamefont
  {Kibble}}, \ and\ \bibinfo {author} {\bibfnamefont {D.~A.}\ \bibnamefont
  {Steer}},\ }\href {\doibase 10.1103/PhysRevD.77.063521} {\bibfield  {journal}
  {\bibinfo  {journal} {Phys. Rev. D}\ }\textbf {\bibinfo {volume} {77}},\
  \bibinfo {pages} {063521} (\bibinfo {year} {2008})},\ \Eprint
  {http://arxiv.org/abs/0712.0808v1} {arXiv:0712.0808v1 [hep-th]} \BibitemShut
  {NoStop}%
%%CITATION = 0712.0808v1;%%
\bibitem [{\citenamefont {Salmi}\ \emph {et~al.}(2008)\citenamefont {Salmi},
  \citenamefont {Ach\'ucarro}, \citenamefont {Copeland}, \citenamefont
  {Kibble}, \citenamefont {de~Putter},\ and\ \citenamefont
  {Steer}}]{SalmiAchucarroCopelandKibblePutterSteer}%
  \BibitemOpen
  \bibfield  {author} {\bibinfo {author} {\bibfnamefont {P.}~\bibnamefont
  {Salmi}}, \bibinfo {author} {\bibfnamefont {A.}~\bibnamefont {Ach\'ucarro}},
  \bibinfo {author} {\bibfnamefont {E.~J.}\ \bibnamefont {Copeland}}, \bibinfo
  {author} {\bibfnamefont {T.~W.~B.}\ \bibnamefont {Kibble}}, \bibinfo {author}
  {\bibfnamefont {R.}~\bibnamefont {de~Putter}}, \ and\ \bibinfo {author}
  {\bibfnamefont {D.~A.}\ \bibnamefont {Steer}},\ }\href {\doibase
  10.1103/PhysRevD.77.041701} {\bibfield  {journal} {\bibinfo  {journal} {Phys.
  Rev. D}\ }\textbf {\bibinfo {volume} {77}},\ \bibinfo {pages} {041701}
  (\bibinfo {year} {2008})},\ \Eprint {http://arxiv.org/abs/0712.1204v2}
  {arXiv:0712.1204v2 [hep-th]} \BibitemShut {NoStop}%
%%CITATION = 0712.1204v2;%%
\bibitem [{\citenamefont {Bevis}\ and\ \citenamefont
  {Saffin}(2008)}]{BevisSaffin}%
  \BibitemOpen
  \bibfield  {author} {\bibinfo {author} {\bibfnamefont {N.}~\bibnamefont
  {Bevis}}\ and\ \bibinfo {author} {\bibfnamefont {P.~M.}\ \bibnamefont
  {Saffin}},\ }\href {\doibase 10.1103/PhysRevD.78.023503} {\bibfield
  {journal} {\bibinfo  {journal} {Phys. Rev. D}\ }\textbf {\bibinfo {volume}
  {78}},\ \bibinfo {pages} {023503} (\bibinfo {year} {2008})},\ \Eprint
  {http://arxiv.org/abs/0804.0200v2} {arXiv:0804.0200v2 [hep-th]} \BibitemShut
  {NoStop}%
%%CITATION = 0804.0200v2;%%
\bibitem [{\citenamefont {Witten}(1985)}]{Witten}%
  \BibitemOpen
  \bibfield  {author} {\bibinfo {author} {\bibfnamefont {E.}~\bibnamefont
  {Witten}},\ }\href {\doibase 10.1016/0550-3213(85)90022-7} {\bibfield
  {journal} {\bibinfo  {journal} {Nuclear Physics B}\ }\textbf {\bibinfo
  {volume} {249}},\ \bibinfo {pages} {557 } (\bibinfo {year}
  {1985})}\BibitemShut {NoStop}%
\bibitem [{\citenamefont {Davis}\ \emph {et~al.}(1997)\citenamefont {Davis},
  \citenamefont {Davis},\ and\ \citenamefont {Trodden}}]{DavisDavisTrodden}%
  \BibitemOpen
  \bibfield  {author} {\bibinfo {author} {\bibfnamefont {S.~C.}\ \bibnamefont
  {Davis}}, \bibinfo {author} {\bibfnamefont {A.~C.}\ \bibnamefont {Davis}}, \
  and\ \bibinfo {author} {\bibfnamefont {M.}~\bibnamefont {Trodden}},\ }\href
  {\doibase 10.1016/S0370-2693(97)00642-4} {\bibfield  {journal} {\bibinfo
  {journal} {Phys.Lett.}\ }\textbf {\bibinfo {volume} {B}},\ \bibinfo {pages}
  {257} (\bibinfo {year} {1997})},\ \Eprint
  {http://arxiv.org/abs/hep-ph/9702360v1} {arXiv:hep-ph/9702360v1 [hep-ph]}
  \BibitemShut {NoStop}%
%%CITATION = hep-ph/9702360v1;%%
\bibitem [{\citenamefont {Davis}\ \emph {et~al.}(1998)\citenamefont {Davis},
  \citenamefont {Davis},\ and\ \citenamefont {Trodden}}]{DavisDavisTrodden2}%
  \BibitemOpen
  \bibfield  {author} {\bibinfo {author} {\bibfnamefont {S.~C.}\ \bibnamefont
  {Davis}}, \bibinfo {author} {\bibfnamefont {A.~C.}\ \bibnamefont {Davis}}, \
  and\ \bibinfo {author} {\bibfnamefont {M.}~\bibnamefont {Trodden}},\ }\href
  {\doibase 10.1103/PhysRevD.57.5184} {\bibfield  {journal} {\bibinfo
  {journal} {Phys.Rev.}\ }\textbf {\bibinfo {volume} {D}},\ \bibinfo {pages}
  {5184} (\bibinfo {year} {1998})},\ \Eprint
  {http://arxiv.org/abs/hep-ph/9711313v1} {arXiv:hep-ph/9711313v1 [hep-ph]}
  \BibitemShut {NoStop}%
%%CITATION = hep-ph/9711313v1;%%
\bibitem [{\citenamefont {Everett}(1988)}]{Everett}%
  \BibitemOpen
  \bibfield  {author} {\bibinfo {author} {\bibfnamefont {A.}~\bibnamefont
  {Everett}},\ }\href {\doibase 10.1103/PhysRevLett.61.1807} {\bibfield
  {journal} {\bibinfo  {journal} {Phys.Rev.Lett.}\ }\textbf {\bibinfo {volume}
  {61}},\ \bibinfo {pages} {1807} (\bibinfo {year} {1988})}\BibitemShut
  {NoStop}%
\bibitem [{\citenamefont {Hindmarsh}\ \emph {et~al.}(2016)\citenamefont
  {Hindmarsh}, \citenamefont {Rummukainen},\ and\ \citenamefont
  {Weir}}]{HindmarshRummukainenWeir}%
  \BibitemOpen
  \bibfield  {author} {\bibinfo {author} {\bibfnamefont {M.}~\bibnamefont
  {Hindmarsh}}, \bibinfo {author} {\bibfnamefont {K.}~\bibnamefont
  {Rummukainen}}, \ and\ \bibinfo {author} {\bibfnamefont {D.~J.}\ \bibnamefont
  {Weir}},\ }\href {\doibase 10.1103/PhysRevLett.117.251601} {\bibfield
  {journal} {\bibinfo  {journal} {Phys.Rev.Lett.}\ }\textbf {\bibinfo {volume}
  {117}},\ \bibinfo {pages} {251601} (\bibinfo {year} {2016})},\ \Eprint
  {http://arxiv.org/abs/arXiv:1607.00764v2} {arXiv:arXiv:1607.00764v2 [hep-th]}
  \BibitemShut {NoStop}%
%%CITATION = arXiv:1607.00764v2;%%
\bibitem [{\citenamefont {Carter}(1989)}]{CARTER1989}%
  \BibitemOpen
  \bibfield  {author} {\bibinfo {author} {\bibfnamefont {B.}~\bibnamefont
  {Carter}},\ }\href {\doibase https://doi.org/10.1016/0370-2693(89)90976-3}
  {\bibfield  {journal} {\bibinfo  {journal} {Physics Letters B}\ }\textbf
  {\bibinfo {volume} {228}},\ \bibinfo {pages} {466 } (\bibinfo {year}
  {1989})}\BibitemShut {NoStop}%
\bibitem [{\citenamefont {Peter}(1992)}]{Peter1992}%
  \BibitemOpen
  \bibfield  {author} {\bibinfo {author} {\bibfnamefont {P.}~\bibnamefont
  {Peter}},\ }\href {\doibase 10.1103/PhysRevD.45.1091} {\bibfield  {journal}
  {\bibinfo  {journal} {Phys. Rev. D}\ }\textbf {\bibinfo {volume} {45}},\
  \bibinfo {pages} {1091} (\bibinfo {year} {1992})}\BibitemShut {NoStop}%
\bibitem [{\citenamefont {Carter}\ and\ \citenamefont
  {Peter}(1995)}]{CarterPeter}%
  \BibitemOpen
  \bibfield  {author} {\bibinfo {author} {\bibfnamefont {B.}~\bibnamefont
  {Carter}}\ and\ \bibinfo {author} {\bibfnamefont {P.}~\bibnamefont {Peter}},\
  }\href {\doibase 10.1103/PhysRevD.52.R1744} {\bibfield  {journal} {\bibinfo
  {journal} {Phys.Rev.}\ }\textbf {\bibinfo {volume} {D}},\ \bibinfo {pages}
  {1744} (\bibinfo {year} {1995})},\ \Eprint
  {http://arxiv.org/abs/hep-ph/9411425v1} {arXiv:hep-ph/9411425v1 [hep-ph]}
  \BibitemShut {NoStop}%
%%CITATION = ARXIV:hep-ph/9411425v1;%%
\bibitem [{\citenamefont {Carter}\ and\ \citenamefont
  {Peter}(1999)}]{CarterPeter2}%
  \BibitemOpen
  \bibfield  {author} {\bibinfo {author} {\bibfnamefont {B.}~\bibnamefont
  {Carter}}\ and\ \bibinfo {author} {\bibfnamefont {P.}~\bibnamefont {Peter}},\
  }\href {\doibase 10.1016/S0370-2693(99)01070-9} {\bibfield  {journal}
  {\bibinfo  {journal} {Phys.Lett.}\ }\textbf {\bibinfo {volume} {B}},\
  \bibinfo {pages} {41} (\bibinfo {year} {1999})},\ \Eprint
  {http://arxiv.org/abs/hep-th/9905025v1} {arXiv:hep-th/9905025v1 [hep-th]}
  \BibitemShut {NoStop}%
%%CITATION = ARXIV:hep-th/9905025v1;%%
\bibitem [{\citenamefont {Carter}(2000)}]{Carter2000}%
  \BibitemOpen
  \bibfield  {author} {\bibinfo {author} {\bibfnamefont {B.}~\bibnamefont
  {Carter}},\ }\href {\doibase
  10.1002/(SICI)1521-3889(200005)9:3/5<247::AID-ANDP247>3.0.CO;2-5} {\bibfield
  {journal} {\bibinfo  {journal} {Ann. Phys.}\ }\textbf {\bibinfo {volume}
  {9}},\ \bibinfo {pages} {247} (\bibinfo {year} {2000})},\ \Eprint
  {http://arxiv.org/abs/hep-th/0002162v1} {arXiv:hep-th/0002162v1 [hep-th]}
  \BibitemShut {NoStop}%
%%CITATION = ARXIV:hep-th/0002162v1;%%
\bibitem [{\citenamefont {Carter}(1990)}]{Carter90}%
  \BibitemOpen
  \bibfield  {author} {\bibinfo {author} {\bibfnamefont {B.}~\bibnamefont
  {Carter}},\ }\href {\doibase 10.1103/PhysRevD.41.3869} {\bibfield  {journal}
  {\bibinfo  {journal} {Phys.Rev.}\ }\textbf {\bibinfo {volume} {D41}},\
  \bibinfo {pages} {3869} (\bibinfo {year} {1990})}\BibitemShut {NoStop}%
\bibitem [{\citenamefont {Vilenkin}(1990)}]{Vilenkin}%
  \BibitemOpen
  \bibfield  {author} {\bibinfo {author} {\bibfnamefont {A.}~\bibnamefont
  {Vilenkin}},\ }\href {\doibase 10.1103/PhysRevD.41.3038} {\bibfield
  {journal} {\bibinfo  {journal} {Phys. Rev.}\ }\textbf {\bibinfo {volume}
  {D41}},\ \bibinfo {pages} {3038} (\bibinfo {year} {1990})}\BibitemShut
  {NoStop}%
\bibitem [{\citenamefont {Carter}(1995)}]{Carter95}%
  \BibitemOpen
  \bibfield  {author} {\bibinfo {author} {\bibfnamefont {B.}~\bibnamefont
  {Carter}},\ }\href {\doibase 10.1103/PhysRevLett.74.3098} {\bibfield
  {journal} {\bibinfo  {journal} {Phys.Rev.Lett.}\ }\textbf {\bibinfo {volume}
  {74}},\ \bibinfo {pages} {3098} (\bibinfo {year} {1995})},\ \Eprint
  {http://arxiv.org/abs/hep-th/9411231v1} {arXiv:hep-th/9411231v1 [hep-th]}
  \BibitemShut {NoStop}%
%%CITATION = ARXIV:hep-th/9411231v1;%%
\bibitem [{\citenamefont {Steer}\ \emph {et~al.}(2018)\citenamefont {Steer},
  \citenamefont {Lilley}, \citenamefont {Yamauchi},\ and\ \citenamefont
  {Hiramatsu}}]{SteerLilleyYamauchiHiramatsu}%
  \BibitemOpen
  \bibfield  {author} {\bibinfo {author} {\bibfnamefont {D.~A.}\ \bibnamefont
  {Steer}}, \bibinfo {author} {\bibfnamefont {M.}~\bibnamefont {Lilley}},
  \bibinfo {author} {\bibfnamefont {D.}~\bibnamefont {Yamauchi}}, \ and\
  \bibinfo {author} {\bibfnamefont {T.}~\bibnamefont {Hiramatsu}},\ }\href
  {\doibase 10.1103/PhysRevD.97.023507} {\bibfield  {journal} {\bibinfo
  {journal} {Phys. Rev. D}\ }\textbf {\bibinfo {volume} {97}},\ \bibinfo
  {pages} {023507} (\bibinfo {year} {2018})},\ \Eprint
  {http://arxiv.org/abs/1710.07475v1} {arXiv:1710.07475v1 [astro-ph.CO]}
  \BibitemShut {NoStop}%
%%CITATION = 1710.07475v1;%%
\bibitem [{\citenamefont {Rybak}\ \emph {et~al.}(2017)\citenamefont {Rybak},
  \citenamefont {Avgoustidis},\ and\ \citenamefont
  {Martins}}]{RybakAvgoustidisMartins}%
  \BibitemOpen
  \bibfield  {author} {\bibinfo {author} {\bibfnamefont {I.~Y.}\ \bibnamefont
  {Rybak}}, \bibinfo {author} {\bibfnamefont {A.}~\bibnamefont {Avgoustidis}},
  \ and\ \bibinfo {author} {\bibfnamefont {C.~J. A.~P.}\ \bibnamefont
  {Martins}},\ }\href {\doibase 10.1103/PhysRevD.96.103535} {\bibfield
  {journal} {\bibinfo  {journal} {Phys. Rev. D}\ }\textbf {\bibinfo {volume}
  {96}},\ \bibinfo {pages} {103535} (\bibinfo {year} {2017})},\ \Eprint
  {http://arxiv.org/abs/1709.01839v2} {arXiv:1709.01839v2 [astro-ph.CO]}
  \BibitemShut {NoStop}%
%%CITATION = 1709.01839v2;%%
\bibitem [{\citenamefont {Blanco-Pillado}\ \emph {et~al.}(2001)\citenamefont
  {Blanco-Pillado}, \citenamefont {Olum},\ and\ \citenamefont
  {Vilenkin}}]{Blanco-PilladoOlumVilenkin}%
  \BibitemOpen
  \bibfield  {author} {\bibinfo {author} {\bibfnamefont {J.~J.}\ \bibnamefont
  {Blanco-Pillado}}, \bibinfo {author} {\bibfnamefont {K.~D.}\ \bibnamefont
  {Olum}}, \ and\ \bibinfo {author} {\bibfnamefont {A.}~\bibnamefont
  {Vilenkin}},\ }\href {\doibase 10.1103/PhysRevD.63.103513} {\bibfield
  {journal} {\bibinfo  {journal} {Phys. Rev. D}\ }\textbf {\bibinfo {volume}
  {63}},\ \bibinfo {pages} {103513} (\bibinfo {year} {2001})},\ \Eprint
  {http://arxiv.org/abs/astro-ph/0004410v3} {arXiv:astro-ph/0004410v3
  [astro-ph]} \BibitemShut {NoStop}%
%%CITATION = astro-ph/0004410v3;%%
\bibitem [{\citenamefont {Davis}\ \emph {et~al.}(2000)\citenamefont {Davis},
  \citenamefont {Kibble}, \citenamefont {Pickles},\ and\ \citenamefont
  {Steer}}]{DavisKibblePicklesSteer}%
  \BibitemOpen
  \bibfield  {author} {\bibinfo {author} {\bibfnamefont {A.~C.}\ \bibnamefont
  {Davis}}, \bibinfo {author} {\bibfnamefont {T.~W.~B.}\ \bibnamefont
  {Kibble}}, \bibinfo {author} {\bibfnamefont {M.}~\bibnamefont {Pickles}}, \
  and\ \bibinfo {author} {\bibfnamefont {D.~A.}\ \bibnamefont {Steer}},\ }\href
  {\doibase 10.1103/PhysRevD.62.083516} {\bibfield  {journal} {\bibinfo
  {journal} {Phys. Rev. D}\ }\textbf {\bibinfo {volume} {62}},\ \bibinfo
  {pages} {083516} (\bibinfo {year} {2000})},\ \Eprint
  {http://arxiv.org/abs/astro-ph/0005514v1} {arXiv:astro-ph/0005514v1
  [astro-ph]} \BibitemShut {NoStop}%
%%CITATION = astro-ph/0005514v1;%%
\bibitem [{\citenamefont {Sharov}(1997)}]{Sharov1997}%
  \BibitemOpen
  \bibfield  {author} {\bibinfo {author} {\bibfnamefont {G.~S.}\ \bibnamefont
  {Sharov}},\ }\href@noop {} {\bibfield  {journal} {\bibinfo  {journal}
  {Theoretical and Mathematical Physics}\ }\textbf {\bibinfo {volume} {113}},\
  \bibinfo {pages} {1263} (\bibinfo {year} {1997})},\ \Eprint
  {http://arxiv.org/abs/hep-th/9808099v1} {arXiv:hep-th/9808099v1 [hep-th]}
  \BibitemShut {NoStop}%
%%CITATION = ARXIV: hep-th/9808099v1;%%
\bibitem [{\citenamefont {'t~Hooft}(2004)}]{Hooft}%
  \BibitemOpen
  \bibfield  {author} {\bibinfo {author} {\bibfnamefont {G.}~\bibnamefont
  {'t~Hooft}},\ }in\ \href@noop {} {\emph {\bibinfo {booktitle} {C04-07-12.5,
  ITP-UU-04-17, SPIN-04-10}}},\ \bibinfo {series} {Hadrons and Strings
  Workshop: A Trento ECT Workshop}, Vol.\ \bibinfo {volume} {5805}\ (\bibinfo
  {address} {Trento, Italy},\ \bibinfo {year} {2004})\ \Eprint
  {http://arxiv.org/abs/hep-th/0408148v1} {arXiv:hep-th/0408148v1 [hep-th]}
  \BibitemShut {NoStop}%
%%CITATION = ARXIV: hep-th/0408148v1;%%
\bibitem [{\citenamefont {Witten}(1996)}]{WITTEN1996}%
  \BibitemOpen
  \bibfield  {author} {\bibinfo {author} {\bibfnamefont {E.}~\bibnamefont
  {Witten}},\ }\href {\doibase https://doi.org/10.1016/0550-3213(95)00610-9}
  {\bibfield  {journal} {\bibinfo  {journal} {Nuclear Physics B}\ }\textbf
  {\bibinfo {volume} {460}},\ \bibinfo {pages} {335 } (\bibinfo {year}
  {1996})},\ \Eprint {http://arxiv.org/abs/hep-th/9510135v2}
  {arXiv:hep-th/9510135v2 [hep-th]} \BibitemShut {NoStop}%
%%CITATION = ARXIV: hep-th/9510135v2;%%
\bibitem [{\citenamefont {Moriarty}\ \emph {et~al.}(1990)\citenamefont
  {Moriarty}, \citenamefont {Myers},\ and\ \citenamefont
  {Rebbi}}]{MoriartyMyersRebbi}%
  \BibitemOpen
  \bibfield  {author} {\bibinfo {author} {\bibfnamefont {K.}~\bibnamefont
  {Moriarty}}, \bibinfo {author} {\bibfnamefont {E.}~\bibnamefont {Myers}}, \
  and\ \bibinfo {author} {\bibfnamefont {C.}~\bibnamefont {Rebbi}},\ }\href
  {\doibase https://doi.org/10.1016/0021-9991(90)90189-8} {\bibfield  {journal}
  {\bibinfo  {journal} {Journal of Computational Physics}\ }\textbf {\bibinfo
  {volume} {88}},\ \bibinfo {pages} {467 } (\bibinfo {year}
  {1990})}\BibitemShut {NoStop}%
\bibitem [{\citenamefont {Laguna}\ and\ \citenamefont
  {Matzner}(1990)}]{LagunaRichard}%
  \BibitemOpen
  \bibfield  {author} {\bibinfo {author} {\bibfnamefont {P.}~\bibnamefont
  {Laguna}}\ and\ \bibinfo {author} {\bibfnamefont {R.~A.}\ \bibnamefont
  {Matzner}},\ }\href {\doibase 10.1103/PhysRevD.41.1751} {\bibfield  {journal}
  {\bibinfo  {journal} {Phys. Rev. D}\ }\textbf {\bibinfo {volume} {41}},\
  \bibinfo {pages} {1751} (\bibinfo {year} {1990})}\BibitemShut {NoStop}%
\end{thebibliography}%
\end{document}